\label{key}\documentclass[10pt,journal,compsoc]{IEEEtran}
\usepackage{amsmath,amssymb,amsfonts}
\usepackage{algorithmic}
\usepackage{graphicx}
\usepackage{textcomp}
\usepackage{xcolor}
\usepackage{amsthm}
\usepackage[ruled, linesnumbered, vlined]{algorithm2e}

\usepackage{pifont} 

\usepackage{textcomp}
\usepackage{multirow}
\usepackage{bm}
\usepackage{booktabs}
\usepackage{epstopdf}
\usepackage{cases}
\usepackage{makecell,multirow,diagbox}
\usepackage{stfloats}
\usepackage{enumitem}[topsep=3pt,itemsep=3pt]
\setlist{nolistsep}
\usepackage{comment}
\usepackage{xcolor}

\newtheorem{theorem}{Theorem}

\newtheorem{definition}{Definition}

\newenvironment{myproof}{\noindent\textit{Proof.}}{\hfill $\blacksquare$}
\usepackage{subfigure} 
\definecolor{color}{rgb}{0.0, 0, 0}
\definecolor{color1}{rgb}{0.0, 0, 0}
\definecolor{b2}{rgb}{0, 0, 0}
\definecolor{b3}{rgb}{0, 0, 0}
\definecolor{b}{rgb}{0, 0, 0}
\definecolor{b1}{rgb}{0, 0, 0}

\allowdisplaybreaks[4]

\def\BibTeX{{\rm B\kern-.05em{\sc i\kern-.025em b}\kern-.08em
		T\kern-.1667em\lower.7ex\hbox{E}\kern-.125emX}}

\ifCLASSOPTIONcompsoc
\usepackage[nocompress]{cite}
\else
\usepackage{cite}
\fi

\begin{document}


\title{Hierarchical Online Optimization Approach for IRS-enabled Low-altitude MEC in Vehicular Networks}

\author{Yixian Wang,
        Geng Sun,~\IEEEmembership{Senior Member,~IEEE},
	Zemin Sun,
        Daxin Tian,~\IEEEmembership{Fellow,~IEEE}, \\ and
        Shiwen Mao,~\IEEEmembership{Fellow,~IEEE}
        \IEEEcompsocitemizethanks{\vspace{-0.5\baselineskip}
        \IEEEcompsocthanksitem Yixian Wang and Zemin Sun are with the College of Computer Science and Technology, Key Laboratory of Symbolic Computation and Knowledge Engineering of Ministry of Education, Jilin University, Changchun 130012, China (e-mail: yixian23@mails.jlu.edu.cn, sunzemin@jlu.edu.cn).
        \IEEEcompsocthanksitem Geng Sun is with the College of Computer Science and Technology, Key Laboratory of Symbolic Computation and Knowledge Engineering of Ministry of Education, Jilin University, Changchun 130012, China, and also with the College of Computing and Data Science, Nanyang Technological University, Singapore 639798 (e-mail: sungeng@jlu.edu.cn).
        \IEEEcompsocthanksitem Daxin Tian is with the School of Transportation Science and Engineering, Beihang University (e-mail: dtian@buaa.edu.cn).
        \IEEEcompsocthanksitem Shiwen Mao is with the Department of Electrical and Computer Engineering, Auburn University, Auburn 36830, USA (e-mail: smao@ieee.org).
        \vspace{0.6\baselineskip}}
        \thanks{{(Corresponding authors: Geng Sun and Zemin Sun.)}}
}

\IEEEtitleabstractindextext{
\begin{abstract}
Low-altitude wireless networks (LAWNs), enabled by uncrewed aerial vehicles (UAVs), are emerging as a key infrastructure to support vehicular networks, where vehicles continuously generate latency-sensitive and computation-intensive applications that require timely computing services. However, the high mobility of vehicles and frequent blockage in complex urban environments lead to highly time-varying air-ground channels, which severely limit the reliable connectivity and consistent computing services. To address this challenge, we propose an intelligent reflecting surface (IRS)-enabled low-altitude multi-access edge computing (MEC) architecture, where an aerial MEC server cooperates with a terrestrial MEC server to provide computing services, while hybrid IRSs (i.e., building-installed and UAV-carried IRSs) are deployed to enhance the air-ground connectivity under blockage. Based on this architecture, we formulate a multi-objective optimization problem (MOOP) to minimize the task completion delay and energy consumption by jointly optimizing task offloading, UAV trajectory control, IRS phase-shift configuration, and computation resource allocation. The considered problem is NP-hard, \textcolor{b}{and thus} we propose a hierarchical online optimization approach (HOOA) to \textcolor{b}{efficiently} solve the problem. Specifically, \textcolor{b}{we reformulate the MOOP} as a Stackelberg game, where MEC servers collectively act as the leader to determine the system-level decisions, while the vehicles act as followers to make individual decisions. At the follower level, we present a many-to-one matching mechanism to generate feasible discrete decisions. At the leader level, we propose a generative diffusion model-enhanced twin delayed deep deterministic policy gradient (GDMTD3) algorithm integrated with a Karush-Kuhn-Tucker (KKT)-based method, which is a deep reinforcement learning (DRL)-based approach, to determine the continuous decisions. \textcolor{b}{Simulation results demonstrate that the proposed HOOA achieves significant improvements, which reduces average task completion delay by 2.5\% and average energy consumption by 3.1\% compared with the best-performing benchmark approach and state-of-the-art DRL algorithm, respectively. Moreover, the proposed HOOA exhibits superior convergence stability while maintaining strong robustness and scalability in dynamic environments.}



\end{abstract}

}

\maketitle
\IEEEdisplaynontitleabstractindextext
\IEEEpeerreviewmaketitle

\section{Introduction}
\label{sec:introduction}
\par Low-altitude wireless networks (LAWNs) are emerging as a key enabler of next-generation connectivity by extending the service capabilities of networks from the ground into the low-altitude airspace~\cite{yuan2025}. By exploiting the mobility of uncrewed aerial vehicles (UAVs) and electric vertical takeoff and landing (eVTOL) platforms, LAWNs can be rapidly deployed to provide supplementary aerial coverage, enhance link availability, and steer capacity toward traffic hotspots~\cite{jin2025}. Compared with conventional terrestrial networks, such aerial augmentation is particularly advantageous in settings with sparse infrastructure, high upgrade costs, or localized service disruption~\cite{zhao2025}, such as intelligent transportation, industrial inspection, and emergency response.


\par Building upon LAWNs, low-altitude multi-access edge computing (MEC) further enhances the role of UAVs by elevating them from pure communication platforms to mobile edge servers, which is capable of executing computation tasks in close proximity to mobile users~\cite{Wu2025a}. Among various application scenarios, vehicular networks represent one of the most demanding and representative use cases of low-altitude MEC. Specifically, vehicular applications such as cooperative perception, high-definition map updating, and real-time navigation continuously generate latency-sensitive and computation-intensive tasks that may exceed the capabilities of onboard processors of vehicles~\cite{Liu2025}. In this case, equipping UAVs with lightweight edge servers enables computing services to be delivered near vehicles, so that alleviating the computation and energy burden on vehicles. However, due to the high mobility of vehicles and UAVs as well as frequent blockage caused by buildings or urban obstacles, air-ground links in low-altitude MEC systems exhibit strong time variations, which severely degrade communication reliability and hinder the consistent provision of computing services. 



\par To mitigate the adverse effects of blockage and enhance the air-ground connectivity in low-altitude MEC, intelligent reflecting surfaces (IRSs) have recently attracted increasing attention as a cost-effective and energy-efficient solution~\cite{Li2025a}. Specifically, an IRS comprises numerous nearly-passive reflecting elements whose complex reflection coefficients are independently configurable, so that the phase of the reflected signal can be controlled~\cite{Ning2025}. By coordinating these elements, the IRS can shape the effective propagation channel, thereby enhancing the power of the received signal and mitigating the link degradation caused by blockage~\cite{xie2025}. Moreover, the hybrid integration of multiple IRSs enables more flexible channel shaping across the network. In addition, phase-controlled passive beamforming avoids active modules such as power amplification and decode-and-forward processing. Therefore, utilizing IRS is more energy-efficient than conventional relaying, as well as incurring lower maintenance overhead and hardware cost~\cite{Ning2025}. 

\par However, fully exploiting the potential of the IRS-enabled low-altitude MEC in vehicular networks still faces several challenges. \textit{First}, the considered system exhibits significant dynamics driven by vehicle and UAV mobility, time-varying air-ground channels, blockage-induced link variations, and stochastic task arrivals, which \textcolor{b}{all make} it challenging to sustain stable and efficient long-term operation\textcolor{b}{~\cite{Dai2024}}. \textit{Second}, from the perspective of problem formulation, existing studies often focus on a single objective, such as service delay or energy consumption. In other words, they do not sufficiently characterize the coordination and conflicts among multiple objectives, so that restricting the attainable system performance\textcolor{b}{~\cite{Dong2024}}. \textit{Third}, from the perspective of decision variables, the joint optimization in this architecture involves both discrete and continuous decisions, and its dimensionality scales with the number of vehicles and IRS elements, which further complicates timely online decision-making\textcolor{b}{~\cite{Jiao2024}}. \textit{Finally}, from the algorithm design perspective, many conventional optimization methods are less effective in dynamic environments\textcolor{b}{~\cite{Shi2023}}, while heuristic algorithms are parameter-sensitive and may yield suboptimal solutions\textcolor{b}{~\cite{Tomar2023}}, and evolutionary algorithms may converge slowly and incur considerable computational overhead\textcolor{b}{~\cite{Li2025b}}. Moreover, conventional deep reinforcement learning (DRL) algorithms still struggle with a strongly coupled and high-dimensional decision space with hybrid actions\textcolor{b}{~\cite{Luong2019}}.

\par To tackle the abovementioned challenges, this work studies multi-objective optimization for the IRS-enabled low-altitude MEC in vehicular networks. The main contributions are summarized below.


\begin{itemize}
    \item \textbf{\textit{System Architecture.}} We consider an IRS-enabled low-altitude MEC architecture for vehicular networks, where an aerial MEC server on the UAV cooperates with a terrestrial MEC server at the base station (BS). Moreover, we introduce a hybrid IRS deployment that combines building-installed and UAV-carried IRSs to enhance air-ground connectivity under blockage and improve service robustness in dynamic environments. To the best of our knowledge, this is the first to investigate hybrid IRS deployment for reliable low-altitude MEC services in vehicular networks. 

    \item \textbf{\textit{Problem Formulation.}} To meet the requirements of latency-sensitive and computation-intensive tasks, we formulate a multi-objective optimization problem (MOOP). Specifically, MOOP jointly optimizes the task offloading, UAV trajectory control, IRS phase-shift configuration, and computation resource allocation to minimize the task completion delay and energy consumption. The formulated MOOP is a mixed-integer nonlinear programming (MINLP) problem, which is generally non-convex and NP-hard.

    \item \textbf{\textit{Approach Design.}} To solve the formulated optimization problem efficiently, we propose a hierarchical online optimization approach (HOOA). Specifically, motivated by the inherent hierarchy in decision-making between the MEC servers and vehicles, the original MOOP is reformulated as a Stackelberg game. At the follower level, a deterministic many-to-one matching mechanism is developed to generate feasible task offloading decisions under server capacity constraints. At the leader level, we propose a generative diffusion model-enhanced twin delayed deep deterministic policy gradient (GDMTD3) algorithm to improve the action representation and exploration for continuous decision-making, while integrating a Karush-Kuhn-Tucker (KKT)-based method to reduce the action dimensionality.


    \item \textbf{\textit{Performance Evaluation.}} Extensive simulation results validate the effectiveness of the proposed HOOA approach. Specifically, the proposed HOOA consistently outperforms the benchmark approaches in terms of average task completion delay, average energy consumption, and average cost of MEC servers. Moreover, compared with state-of-the-art DRL algorithms, HOOA exhibits faster convergence and stronger learning stability, achieving higher and smoother rewards during training together with more stable delay and energy trends. Furthermore, the hyper-parameter sensitivity analysis corroborates the effectiveness of the adopted settings. Finally, evaluations under different task sizes and varying numbers of vehicles demonstrate that HOOA achieves good robustness and scalability in dynamic vehicular network scenarios.
\end{itemize}

\par The remainder of the paper is structured as follows. Section~\ref{sec:related_work} reviews related work. Section~\ref{sec:system_model} introduces the system model. Building on this model, Section~\ref{sec:problem_formulation_and_analysis} formulates and analyzes the optimization problem. Section~\ref{sec:HOOA} describes the proposed HOOA approach in detail. Section~\ref{sec:simulation_results} presents simulation results. Finally, Section~\ref{sec:conclusion} summarizes the paper.
\vspace{-1.3em}
\section{Related Work}
\label{sec:related_work}

\par In this section, we review related work on low-altitude MEC architecture, joint optimization problem formulation, and optimization approach.
\vspace{-1.2em}
\subsection{Low-altitude MEC Architecture}
\label{sec:low_altitude_mec_architecture}
\vspace{-0.2em}
\par Conventional terrestrial MEC is vulnerable to service congestion and performance instability under concentrated vehicular workloads due to its reliance on fixed edge nodes such as roadside units and BSs, which has driven extensive studies on low-altitude MEC. For example, Li \textit{et al.}~\cite{Li2025} introduced UAVs as the mobile aerial edge nodes to relieve the workload of terrestrial edge servers and improve the service availability in demand hotspots. Moreover, Zhang \textit{et al.}~\cite{Zhang2024} incorporated computation-capable UAVs into networks with multiple vehicles and terrestrial edge servers, so that the UAVs can simultaneously support aerial relaying and task execution. However, such architectures remain fundamentally constrained by onboard energy and coverage range in large-scale deployments. In addition, the uncertain propagation conditions in complex urban environments further degrade the stability of air-ground connectivity.

\par Leveraging the capability of IRSs to programmably reshape radio propagation, recent studies have integrated IRSs with low-altitude MEC to improve link robustness and service availability. For instance, Wu \textit{et al.}~\cite{Wu2025} developed an upgraded MEC system that integrates IRSs and UAVs into a terahertz communication network to extend effective coverage and mitigate blockage effects. Furthermore, Gao \textit{et al.}~\cite{Gao2025} proposed a multi-IRS-assisted low-altitude MEC architecture, where distributed IRSs cooperatively enhance the air-ground link confidentiality to support secure task offloading. However, most of these studies focus on building-installed IRSs, which limits their adaptability to rapidly changing user locations and channel conditions in dynamic low-altitude MEC environments, thus compromising the stability of the IRS-enabled link quality.

\par To overcome the abovementioned challenges, recent studies have further explored UAV-carried IRS-enabled MEC. For example, Liao \textit{et al.}~\cite{Liao2025} presented a reconfigurable intelligent surface (RIS)-assisted UAV-unmanned surface vehicle (USV) cooperative MEC architecture, which supports bidirectional tasks of USVs under hard time-window constraints. In addition, Jiang \textit{et al.}~\cite{Jiang2024} investigated a multiple-aerial IRS-assisted MEC architecture, in which aerial IRSs are deployed to enable timely and reliable task offloading from devices to an edge server in poor offloading environments. However, the IRS component in these studies mainly serves as a communication enhancer for coverage and reliability, while computation is still executed at terrestrial edge servers or end devices, thereby leaving joint communication and computation design insufficiently explored in dynamic deployments.

\par In summary, existing low-altitude MEC architectures remain challenged in sustaining service continuity under mobility and blockage. Meanwhile, IRS-enabled low-altitude MEC architectures typically rely on either building-installed IRS deployments with limited adaptivity or UAV-carried IRS deployments that primarily enhance wireless links without strong coordination between communication and computation. Therefore, this work considers a hybrid IRS-enabled low-altitude MEC architecture in which aerial and terrestrial MEC servers collaboratively provide edge computing services, while a hybrid IRS deployment of building-installed and UAV-carried IRSs improves robustness and flexibility in dynamic environments.
\vspace{-1em}
\subsection{Formulation of Joint Optimization Problems}
\label{sec:formulation_joint_optimization_problems}
\par Formulating a joint optimization problem is essential to assess the system-level performance of the considered IRS-enabled low-altitude MEC architecture for vehicular networks. Existing studies on IRS-enabled low-altitude MEC have investigated various design objectives, such as service delay and energy consumption. For example, Zhou \textit{et al.}~\cite{Zhou2025} considered an IRS-UAV-assisted wireless power transfer-MEC system and formulated a latency minimization problem under an energy-consumption constraint. Besides, Alshahrani~\cite{Alshahrani2025} presented an optimization framework for a BS-hosted MEC system aided by a UAV-equipped RIS to minimize task execution latency. Moreover, for a RIS-assisted wireless-powered MEC system where a UAV-mounted cloudlet serves multiple user equipment, Kim \textit{et al.}~\cite{Kim2023} developed an energy-consumption minimization problem. However, these studies typically optimize only one single aspect of system performance, which can lead to suboptimal designs when tasks simultaneously require low service delay and low energy consumption.

\par In addition to the optimization objective, system performance is also strongly influenced by how the decision variables are jointly optimized. Previous studies on IRS-enabled low-altitude MEC have explored the optimization of various decision variables, such as task offloading, UAV trajectory control, and IRS phase-shift configuration. For example, Zeng \textit{et al.}~\cite{Zeng2025} proposed a simultaneously transmitting-RIS (STAR-RIS)-assisted low-altitude MEC system by jointly optimizing time-slot allocation, STAR-RIS coefficient matrices, and UAV trajectory. Moreover, Michailidis \textit{et al.}~\cite{Michailidis2021} jointly optimized time-slot scheduling and task allocation subject to transmit power constraints for the dual-RIS-assisted UAV-aided internet of vehicles (IoV) offloading architecture. Furthermore, Li \textit{et al.}~\cite{Li2024} jointly optimized the task partition parameters and the transmit power of all mobile users, together with the IRS reflection coefficient matrix and UAV trajectory for an IRS low-altitude MEC framework. However, the abovementioned studies do not explicitly optimize the computation resource allocation, which is essential for coordinating MEC server execution and meeting stringent delay requirements under time-varying offloading demands.

\par This work differs from the studies above in terms of the optimization objectives and decision variables. Specifically, our optimization objectives account for the task completion delay and energy consumption to capture their inherent trade-off in dynamic vehicular environments. In addition, we jointly optimize a more comprehensive set of decision variables, including task offloading, UAV trajectory control, IRS phase-shift configuration, and computation resource allocation for the considered IRS-enabled low-altitude MEC in vehicular networks.
\vspace{-1.2em}
\subsection{Optimization Approaches}
\label{sec:optimization_approach}

\par To tackle the challenging optimization problems, relevant studies have proposed efficient approaches based on optimization theory, evolutionary algorithms, and heuristic algorithms. For instance, Xiao \textit{et al.}~\cite{Xiao2025} employed a block coordinate descent iterative framework integrating the Dinkelbach method and successive convex approximation (SCA) to handle a strongly coupled non-convex formulation in a STAR-RIS-enhanced low-altitude MEC system. Moreover, Zhang \textit{et al.}~\cite{Zhang2024a} considered a UAV-deployed RIS-aided MEC system with non-orthogonal multiple access, where the joint design was solved by using complex circle manifold optimization and a genetic algorithm. Furthermore, Liao \textit{et al.}~\cite{Liao2024} presented a heuristic iterative scheme for RIS-assisted cooperative UAV-USV MEC, in which a modified alternating direction method of multipliers algorithm, enhanced simulated annealing, and an SCA-based routine were adopted. Despite their effectiveness, many traditional optimization methods become significantly limited in highly dynamic environments. In addition, heuristic algorithms can be highly sensitive to parameter tuning and are prone to being trapped in suboptimal solutions, whereas evolutionary algorithms typically require many iterations, thus resulting in slow convergence and considerable computational overhead.

\par Given the high dynamics and complex coupling in IRS-enabled low-altitude MEC systems, DRL has been increasingly adopted as a model-free approach for adaptive decision-making. For example, Wu \textit{et al.}~\cite{Wu2024} proposed an energy efficiency maximization scheme under energy constraints based on double deep Q-network (DDQN). Moreover, Chen \textit{et al.}~\cite{Chen2025a} adopted a proximal policy optimization (PPO)-based algorithm to iteratively learn a policy through continuous interaction with the environment to maximize system energy efficiency. Furthermore, Pang \textit{et al.}~\cite{Pang2025} incorporated a DRL-based softmax deep double deterministic policy gradients algorithm to optimize the system with the objective of enhancing energy harvesting performance. Nevertheless, conventional DRL algorithms still struggle to cope with a strongly coupled and high-dimensional decision space with variable and hybrid actions.

\par To address these challenges, this paper proposes HOOA, which reformulates the MOOP into a Stackelberg game. Under this framework, a many-to-one matching mechanism is employed to generate feasible discrete decisions for vehicles. Moreover, we propose the GDMTD3 algorithm to enhance action representation and exploration for continuous decisions, and further integrate a KKT-based method to reduce action dimensionality, thereby improving decision efficiency while achieving better overall performance.

%
%
\vspace{-1em}
\section{System Model}
\label{sec:system_model}
\par In this section, we introduce the considered IRS-enabled low-altitude MEC architecture for vehicular networks. Moreover, we present the basic models, the communication model, and the computation model.
\vspace{-1em}
\subsection{System Overview}
\label{sec:system_overview}

\subsubsection{System Architecture}
\label{sec:system_architecture}
\par
\begin{figure}[!t]
    \centering
    \includegraphics[width=3.25in]{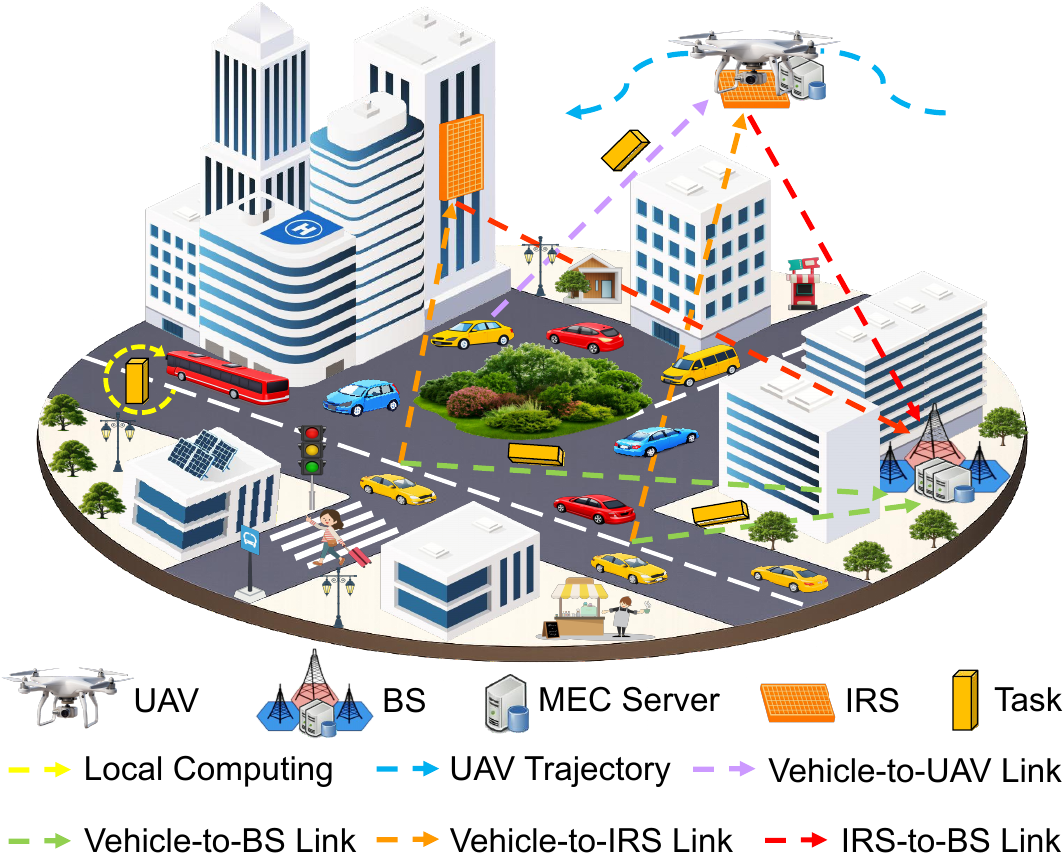}
    \vspace{-1em}
    \caption{IRS-enabled low-altitude MEC architecture for vehicular networks, where vehicles offload tasks to the UAV or BS, and hybrid IRSs enhance the communication quality for task offloading to the BS.}
    \label{fig_System_Model}
\end{figure}

 As shown in Fig.~\ref{fig_System_Model}, we consider an IRS-enabled low-altitude MEC architecture for vehicular networks, which includes a set of vehicles denoted by $\mathcal{I}=\{1,2,\ldots,I\}$, where each vehicle generates computation tasks such as intelligent parking and online navigation. Moreover, the architecture comprises a UAV $u$ equipped with an aerial MEC server that provides flexible computing services to vehicles within its service range, and a BS $b$ equipped with a terrestrial MEC server that offers reliable computing services. Notably, the aerial MEC server and terrestrial MEC server are collectively referred to as MEC servers, indexed by $j\in\{u,b\}$. Furthermore, we assume that the presence of obstacles limits the direct communication link between the vehicles and the BS~\cite{Jameel2019}. In this case, we adopt a hybrid deployment of a set of IRSs denoted by $\mathcal{K}=\{1,2,\ldots,K\}$, including both building-installed and UAV-carried IRSs. Specifically, each IRS $k\in\mathcal{K}$ consists of $L$ reflective elements denoted by $\mathcal{L}=\{1,2,\ldots,L\}$, and it is equipped with a controller to adjust the phase shift of each reflecting element. In addition, for ease of exposition, the continuous system \textcolor{b}{time with duration} $T$ is discretized into a set of time slots denoted by $\mathcal{N}=\{1,2,\ldots,N\}$, where each slot duration $\delta_t=T/N$ is selected to be sufficiently small so that the system dynamics can be considered constant within each time slot and only vary across time slots.
\vspace{-1em}

\subsubsection{Basic Models}
\label{sec:basic_models}
\par The basic models of the system are illustrated below.
\par\textit{\textbf{Vehicle Mobility Model.}} The horizontal coordinate of each vehicle $i\in\mathcal{I}$ in time slot $n$ is denoted by $\mathbf{q}_i(n)=\big[x_i(n), y_i(n)\big]^{\mathrm{T}}$. Moreover, we assume that the movement of vehicles follows the Gauss-Markov mobility model~\cite{Batabyal2015}. Thus, the position of vehicle $i$ can be updated as
\begin{equation}
\label{eq:vehicle_position}
    \mathbf{q}_i(n+1) = \mathbf{q}_i(n) + \mathbf{v}_{i}(n)\delta_t.
\end{equation}


\par\textit{\textbf{UAV Mobility Model.}} With a fixed altitude $H$, the instantaneous horizontal coordinate of UAV $u$ in time slot $n$ is denoted by
$\mathbf{q}_u(n)=\big[x_u(n), y_u(n)\big]^{\mathrm{T}}$. Hence, the position of UAV $u$ can be updated as
\begin{subequations}
\label{eq:uav_position}
\begin{align}
    x_u(n+1) &= x_u(n) + \delta_t\,v_u(n)\cos\big(\varphi_u(n)\big), \label{eq:uav_position_x}\\
    y_u(n+1) &= y_u(n) + \delta_t\,v_u(n)\sin\big(\varphi_u(n)\big), \label{eq:uav_position_y}
\end{align}
\end{subequations}
\noindent where $\varphi_u(n)\in[-\pi,\pi)$ denotes the heading angle of UAV $u$, $v_u(n)\in[0,v_u^{\max}]$ represents the speed of UAV $u$, and $v_u^{\max}$ is the maximum allowable speed. In addition, the position of the UAV is subject to the following physical constraints
\begin{subequations}
\label{eq:uav_area_constraints}
\begin{alignat}{2}
    & 0 \leq x_u(n) \leq x^{\max}, && \ \forall n \in \mathcal{N}, \label{eq:uav_area_x}\\
    & 0 \leq y_u(n) \leq y^{\max}, && \ \forall n \in \mathcal{N}, \label{eq:uav_area_y}
\end{alignat}
\end{subequations}
\noindent where constraints \eqref{eq:uav_area_x} and \eqref{eq:uav_area_y} specify the feasible horizontal flight region of the UAV.

\par\textit{\textbf{Vehicle Model.}} Similar to~\cite{Huang2023}, we assume that each vehicle can generate multiple computation tasks during the system timeline, with one task generated in each time slot. Accordingly, each vehicle $i\in\mathcal{I}$ is described by $\langle F_i^{\max}, \zeta_i(n) \rangle$, where $F_i^{\max}$ represents the total computing capability of vehicle $i$ and $\zeta_i(n)$ refers to the task generated by vehicle $i$ in time slot $n$. To be more specific, the generated task $\zeta_i(n)$ can be characterized by the tuple $\langle D_i(n), G_i(n), T_i^{\max}(n) \rangle$, wherein $D_i(n)$ is the task size (in bits), $G_i(n)$ represents the computation intensity of the task (cycles/bit), and $T_i^{\max}(n)$ denotes the deadline of the task. Furthermore, due to the limited onboard computing resources, we consider that each vehicle is equipped with a single CPU core~\cite{Dai2019}.

\par\textit{\textbf{Server Model.}} Each MEC server $j\in\{u,b\}$ is equipped with a multi-core CPU so that multiple computation tasks can be processed in parallel~\cite{Dai2019}. Consequently, each MEC server $j\in\{u,b\}$ is characterized by $\langle F_j^{\max}, m_j^{\text{core}} \rangle$, where $F_j^{\max}$ represents the total computing capability of MEC server $j$, and $m_j^{\text{core}}$ indicates the number of CPU cores available at MEC server $j$.
\vspace{-1em}
\subsection{Communication Model}
\label{sec:communication_model}

\par We consider two types of communication links, namely the direct link (i.e., vehicle-to-UAV link) and reflected link assisted by the hybrid IRS deployment (i.e., vehicle-to-BS link, vehicle-to-IRS link, and IRS-to-BS link). In addition, we adopt orthogonal frequency division multiple access (OFDMA)~\cite{Han2007} to support simultaneous uplink transmissions from multiple vehicles and enhance transmission reliability by mitigating mutual interference from concurrent transmissions. Specifically, the communication links mentioned above are described as follows.
\vspace{-0.8em}
\subsubsection{Direct Link}
\label{sec: Direct_Link}
\par Due to the presence of LoS and non-line-of-sight (NLoS) components, the channel between vehicle $i$ and UAV $u$ in time slot $n$ follows Rician fading~\cite{You2019}, which is given as
\begin{equation}
\label{eq:channel_vehicle_uav}
    h_{i,u}(n)
    = \sqrt{\rho d_{i,u}^{-\alpha_{i,u}}(n)}
    \big(\sqrt{\tfrac{\gamma^{\mathrm{rf}}}{1+\gamma^{\mathrm{rf}}}}\,h_{i,u}^{\mathrm{LoS}}(n)
    + \sqrt{\tfrac{1}{1+\gamma^{\mathrm{rf}}}}\,h_{i,u}^{\mathrm{NLoS}}(n)\big),
\end{equation}
\noindent where $\alpha_{i,u}$ denotes the associated path loss exponent, $\gamma^{\mathrm{rf}}$ represents the Rician factor, and $d_{i,u}(n)$ is the distance between vehicle $i$ and UAV $u$ in time slot $n$. Moreover, the LoS component and NLoS component are denoted by $h_{i,u}^{\mathrm{LoS}}(n)$ and $h_{i,u}^{\mathrm{NLoS}}(n)$, respectively. Therefore, in time slot $n$, the data transmission rate between vehicle $i$ and UAV $u$ is expressed as
\begin{equation}
\label{eq:rate_vehicle_uav}
    R_{i,u}(n)
    = B_{i,u}(n)\log_2\!\big(1+p_i^{\text{tr}}(n)\big|h_{i,u}(n)\big|^2/\sigma^2\big),
\end{equation}
\noindent where $B_{i,u}(n)$ represents the bandwidth allocated to vehicle $i$ for transmission to UAV $u$ in time slot $n$, $p_i^{\text{tr}}(n)$ denotes the transmit power of vehicle $i$, and $\sigma^2$ is the noise power.
\vspace{-0.8em}
\subsubsection{Reflected Link}
\label{sec: Reflected_Link}
\par The channel models for vehicle-to-BS link, vehicle-to-IRS link, and IRS-to-BS link are detailed as follows.

\par\textit{\textbf{Vehicle-to-BS Link.}} Considering the complex propagation environment with obstacles between the vehicles and the BS, we model the channel from vehicle $i$ to BS $b$ in time slot $n$ as Rayleigh fading~\cite{Pang2022}, which is given as
\begin{equation}
\label{eq:channel_vehicle_bs}
    h_{i,b}(n)
    = \sqrt{\rho d_{i,b}^{-\alpha_{i,b}}(n)}\,\tilde{h}_{i,b}(n),
\end{equation}

\noindent where $\alpha_{i,b}$ denotes the path loss exponent, $\rho$ represents the path loss at the reference distance of 1 m, $d_{i,b}(n)$ is the distance between vehicle $i$ and BS $b$ in time slot $n$, and $\tilde{h}_{i,b}(n)$ follows a zero-mean and unit-variance complex Gaussian distribution.

\par\textit{\textbf{Vehicle-to-IRS Link.}} Each IRS $k\in\mathcal{K}$ is placed or maneuvered to ensure a dominant LoS component between vehicle $i$ and IRS $k$. Hence, the channel $\boldsymbol{h}_{i,k}(n)\in\mathbb{C}^{1\times L}$ from vehicle $i$ to IRS $k$ in time slot $n$ follows Rician fading~\cite{Chen2021} as follows
\begin{equation}
\label{eq:channel_vehicle_irs}
    \boldsymbol{h}_{i,k}(n)
    = \sqrt{\rho d_{i,k}^{-\alpha_{i,k}}(n)}
    \big(\sqrt{\tfrac{\gamma^{\mathrm{rf}}}{1+\gamma^{\mathrm{rf}}}}\,\boldsymbol{h}_{i,k}^{\mathrm{LoS}}(n)
    + \sqrt{\tfrac{1}{1+\gamma^{\mathrm{rf}}}}\,\boldsymbol{h}_{i,k}^{\mathrm{NLoS}}(n)\big),
\end{equation}
\noindent where $\alpha_{i,k}$ denotes the associated path loss exponent, and $d_{i,k}(n)$ is the distance between vehicle $i$ and the reference point of IRS $k$. Moreover, the LoS component and NLoS component are represented by $\boldsymbol{h}_{i,k}^{\mathrm{LoS}}(n)$ and $\boldsymbol{h}_{i,k}^{\mathrm{NLoS}}(n)$, respectively.

\par\textit{\textbf{IRS-to-BS Link.}} Similar to~\cite{Chen2021}, the channel from IRS $k$ to BS $b$ in time slot $n$ also follows Rician fading. Moreover, $\boldsymbol{h}_{k,b}(n)\in\mathbb{C}^{L\times 1}$ can be modeled in the same form as Eq.~\eqref{eq:channel_vehicle_irs}, and it is omitted here due to space constraints.

\par In addition, the reflection-coefficient matrix of IRS $k$ in time slot $n$ is given as~\cite{Wu2021}
\begin{equation}
\label{eq:irs_phase_matrix}
    \boldsymbol{\Theta}_{k}(n)
    = \operatorname{diag}\big(\beta e^{\iota \theta_{k,1}(n)}, \ldots, \beta e^{\iota \theta_{k,L}(n)}\big),
\end{equation}
\noindent where $\beta$ is set to 1 and $\iota$ is the imaginary unit ($\iota=\sqrt{-1}$).

\par Based on the channel model above, the corresponding signal-to-noise ratio (SNR) at BS $b$ for vehicle $i$ in time slot $n$ is given as
\begin{equation}
\label{eq:sinr_vehicle_bs}
    \delta_{i,b}(n)
    = p_i^{\text{tr}}(n)\big|
    h_{i,b}(n)
    + \sum_{k\in\mathcal{K}}
    \boldsymbol{h}_{i,k}(n)\boldsymbol{\Theta}_{k}(n)\boldsymbol{h}_{k,b}(n)
    \big|^2/\sigma^2.
\end{equation}

\par Consequently, in time slot $n$, the data transmission rate between vehicle $i$ and BS $b$ can be expressed as
\begin{equation}
\label{eq:rate_vehicle_bs}
    R_{i,b}(n)
    = B_{i,b}(n)\log_2\big(1+\delta_{i,b}(n)\big),
\end{equation}
\noindent where $B_{i,b}(n)$ is the bandwidth allocated to vehicle $i$ for transmission to BS $b$ in time slot $n$.
\vspace{-1em}
\subsection{Computation Model}
\label{sec:computation_model}

\subsubsection{Service Delay}
\label{sec:service_delay}

\par The service delay for task completion is determined by the offloading decision. Specifically, the task $\zeta_i(n)$ can be processed either locally on vehicle $i$ or remotely on MEC server $j \in \{u,b\}$ (i.e., directly offloaded to the UAV $u$, or transmitted to the BS $b$ via the hybrid IRS deployment). To this end, we introduce binary offloading indicators $O_i^{a}(n)\in\{0,1\}$, $a\in\mathcal{A}=\{i\}\cup\{j\mid j\in\{u,b\}\}$, to represent the offloading decision of vehicle $i$ in time slot $n$. Note that for edge computing, we ignore the result feedback delay since the results of most mobile applications are typically much smaller than the input data~\cite{Mao2017}.

\par \textbf{\textit{Local Computing.}} The service delay of vehicle $i$ to process task $\zeta_i(n)$ locally in time slot $n$ is given as
\begin{equation}
\label{eq:delay_local}
    T_i^{i}(n) = D_i(n)G_i(n)/F_i^{\max}.
\end{equation}

\par \textbf{\textit{Edge Computing.}} When task $\zeta_i(n)$ is processed by MEC server $j\in\{u,b\}$, the offloading service delay consists of the transmission delay and computation delay, which is expressed as
\begin{equation}
\label{eq:delay_edge}
    T_i^{j}(n)
    = \underbrace{D_i(n)/R_{i,j}(n)}_{\text{Transmission}}
    + \underbrace{D_i(n)G_i(n)/f_{j,i}(n)}_{\text{Computation}},
\end{equation}
\noindent where $f_{j,i}(n)$ is the computation resource allocated by MEC server $j$ to task $\zeta_i(n)$ in time slot $n$.

\par According to Eqs.~\eqref{eq:delay_local} and~\eqref{eq:delay_edge}, the total task completion delay across $N$ time slots is written as
\begin{align}
\label{eq:delay_total}
    T^{\text{total}}
    &= \sum_{i\in \mathcal{I}}\sum_{n\in \mathcal{N}}
    \big( O_{i}^{i}(n)\, T_i^{i}(n)
    + \sum_{j \in \{u,b\}} O_{i}^{j}(n)\, T_i^{j}(n) \big).
\end{align}

\subsubsection{Energy Consumption}
\label{sec:energy_consumption}
\par Processing task $\zeta_i(n)$ may impose additional costs on vehicles or MEC servers.

\par \textbf{\textit{Local Computing.}} The energy consumption of vehicle $i$ to process task $\zeta_i(n)$ locally in time slot $n$ is given as
\begin{equation}
\label{eq:energy_local}
    E_{i}^{i}(n)
    = \kappa_i \big(F_i^{\max}\big)^2 D_i(n)G_i(n),
\end{equation}
\noindent where $\kappa_i \geq 0$ denotes the effective switched capacitance of the CPU in vehicle $i$~\cite{Chen2025}.

\par \textbf{\textit{Edge Computing.}} When task $\zeta_i(n)$ is offloaded to MEC server $j \in \{u,b\}$, the offloading energy consumption consists of the transmission energy and computation energy, which is expressed as
\begin{equation}
\label{eq:energy_edge}
    E_i^{j}(n)
    = \underbrace{p_i^{\text{tr}}(n)D_i(n)/R_{i,j}(n)}_{\text{Transmission}}
    + \underbrace{\varpi_j D_i(n)G_i(n)}_{\text{Computation}},
\end{equation}
\noindent where $\varpi_j$ represents the effective computation energy coefficient of MEC server $j$~\cite{Jiang2023}.

\par \textbf{\textit{UAV Flight Energy.}} Similar to~\cite{Zeng2019a}, the flight energy consumption of the UAV $u$ in time slot $n$ is given as
\begin{equation}
\label{eq:energy_uav_flight}
\begin{aligned}
    E_u^{\text{fly}}(n)
    &= \delta_t\big(
    \eta_1 \big(1 + 3v_u^2(n)/U_{\text{tip}}^2\big)
    + \eta_4 v_u^3(n) \\
    &\quad + \eta_2 \sqrt{
    \sqrt{\eta_3 + v_u^4(n)/4}
    - v_u^2(n)/2
    }\big),
\end{aligned}
\end{equation}
\noindent where $U_{\text{tip}}$ is the blade tip speed of the rotor. Moreover, $\eta_1, \eta_2, \eta_3,$ and $\eta_4$ are constants determined by the aerodynamic properties.

\par According to Eqs.~\eqref{eq:energy_local},~\eqref{eq:energy_edge}, and~\eqref{eq:energy_uav_flight}, the total energy consumption across $N$ time slots is written as
\begin{align}
\label{eq:energy_total}
    E^{\text{total}}
    &= \sum_{i\in \mathcal{I}} \sum_{n\in \mathcal{N}}
    \big(
    O_{i}^{i}(n)\, E_i^{i}(n)
    + \sum_{j \in \{u,b\}} O_{i}^{j}(n)\, E_i^{j}(n)
    \big) \nonumber\\
    &\quad + \sum_{n\in \mathcal{N}} E_u^{\text{fly}}(n).
\end{align}

%
%
\vspace{-1em}
\section{Problem Formulation and Analysis}
\label{sec:problem_formulation_and_analysis}

\subsection{Problem Formulation}
\label{sec:problem_formulation}
\par This work aims to minimize the total task completion delay and total energy consumption of the system by jointly optimizing task offloading $\mathbf{O} = \{ O_{i}^{a}(n) \}_{i \in \mathcal{I}, a \in \mathcal{A}, n \in \mathcal{N}}$, UAV trajectory control $\mathbf{Q}=\{\mathbf{q}_u(n)\}_{n \in \mathcal{N}}$, IRS phase-shift configuration $\boldsymbol{\theta} = \{\theta_{k,l}(n)\}_{k \in \mathcal{K}, l \in \mathcal{L}, n \in \mathcal{N}}$, and computation resource allocation $\mathbf{F} = \{f_{j,i}(n)\}_{i \in \mathcal{I}, j \in \{u,b\}, n \in \mathcal{N}}$. Therefore, the MOOP can be formulated as
\begin{subequations}
\label{eq:optimization_problem}
\begin{align}
    \mathbf{P}: \quad
    & \min _{\mathbf{O},\mathbf{Q},\boldsymbol{\theta},\mathbf{F}} \
    \big\{ T^{\text{total}},  E^{\text{total}} \big\}
    \label{eq:optimization_objective} \\
    \text{s.t.}\quad
    & O_i^a(n)\in\{0,1\}, \forall i\in \mathcal{I}, \ a\in \mathcal{A}, \ n \in \mathcal{N},
    \label{eq:constraint_offload_binary} \\
    & \sum_{a\in \mathcal{A}} O_i^a(n)= 1, \forall i\in \mathcal{I}, \ n \in \mathcal{N},
    \label{eq:constraint_offload_unique} \\
    & O_i^a(n)\,T_i^a(n)\leq T_i^{\max}(n), \forall i\in \mathcal{I}, \ a\in \mathcal{A}, \ n \in \mathcal{N},
    \label{eq:constraint_deadline} \\
    & \sum_{i\in \mathcal{I}} O_i^{j}(n) \leq m_j^{\text{core}}, \forall j \in \{u,b\}, \ n \in \mathcal{N},
    \label{eq:constraint_core_limit} \\
    & 0 \leq \theta_{k,l}(n)<2 \pi, \forall k \in \mathcal{K}, \ l \in \mathcal{L}, \ n \in \mathcal{N},
    \label{eq:constraint_phase_bounds} \\
    & \sum_{i\in\mathcal{I}} O_i^{j}(n)\, f_{j,i}(n) \le F_j^{\max}, \forall j \in \{u,b\}, \ n \in \mathcal{N},
    \label{eq:constraint_cpu_sum} \\
    & \text{\eqref{eq:vehicle_position}-\eqref{eq:uav_area_constraints}.}
    \label{eq:constraint_mobility_models}
\end{align}
\end{subequations}
\noindent Constraints \eqref{eq:constraint_offload_binary} and \eqref{eq:constraint_offload_unique} impose a choice between local computing and offloading for each vehicle. Constraint \eqref{eq:constraint_deadline} ensures that each task is completed within its deadline. Constraint \eqref{eq:constraint_core_limit} restricts MEC server $j$ to no more than $m_j^{\text{core}}$ tasks in each time slot. Moreover, constraint \eqref{eq:constraint_phase_bounds} specifies that the phase shift of each IRS element $l$ ranges within $[0,2\pi)$. Constraint \eqref{eq:constraint_cpu_sum} ensures that the total computation resources allocated by each MEC server to the offloaded tasks do not exceed its maximum computing capacity. In addition, constraint \eqref{eq:constraint_mobility_models} enforces the mobility models of the vehicles and the UAV.

\vspace{-1em}
\subsection{Problem Analysis}
\label{sec:problem_analysis}

\par It is challenging to solve the formulated MOOP directly for several reasons, which can be summarized as follows.

\begin{itemize}

    \item \textit{Multi-objective trade-offs across heterogeneous decision entities.} The formulated MOOP involves inherently conflicting objectives. On the one hand, minimizing the total task completion delay requires more aggressive task offloading, frequent UAV repositioning, and effective IRS phase-shift configurations, which may increase communication workload or UAV flight energy consumption\textcolor{b}{~\cite{Sun2024}}. Conversely, energy-aware decisions often lead to increased task latency. More importantly, these conflicting objectives are associated with different decision entities operating under asymmetric information and decision privileges, which makes it difficult to coordinate the trade-off within a centralized optimization framework.
    
    \item \textit{NP-hard and non-convex optimization.} MOOP involves continuous variables (i.e., UAV trajectory control $\mathbf{Q}$, IRS phase-shift configuration $\boldsymbol{\theta}$, and computation resource allocation $\mathbf{F}$) and discrete variables (i.e., task offloading $\mathbf{O}$). Moreover, the objectives and constraints are nonlinear due to the tight coupling between the communication and computation processes. Consequently, MOOP is an MINLP problem~\cite{Santis2020}, which is generally non-convex and NP-hard.
    
    \item \textit{Inter-slot coupling and long-horizon dependence.} The decision variables are indexed over time slots, and the UAV mobility constraints impose sequential dependence across slots. Meanwhile, random task arrivals and vehicle mobility jointly make the workload and link conditions time-varying, such that the decisions cannot be optimized independently on a per-slot basis\textcolor{b}{~\cite{Sun2025}}. As a result, MOOP becomes a long-horizon coupled optimization, in which per-slot optimization strategies are insufficient to achieve stable long-term system performance.
    
\end{itemize}

%
%
\vspace{-1em}
\section{The Proposed HOOA}
\label{sec:HOOA}
\par
\begin{figure*}[!hbt] 
	\centering
	\includegraphics[width =6in]{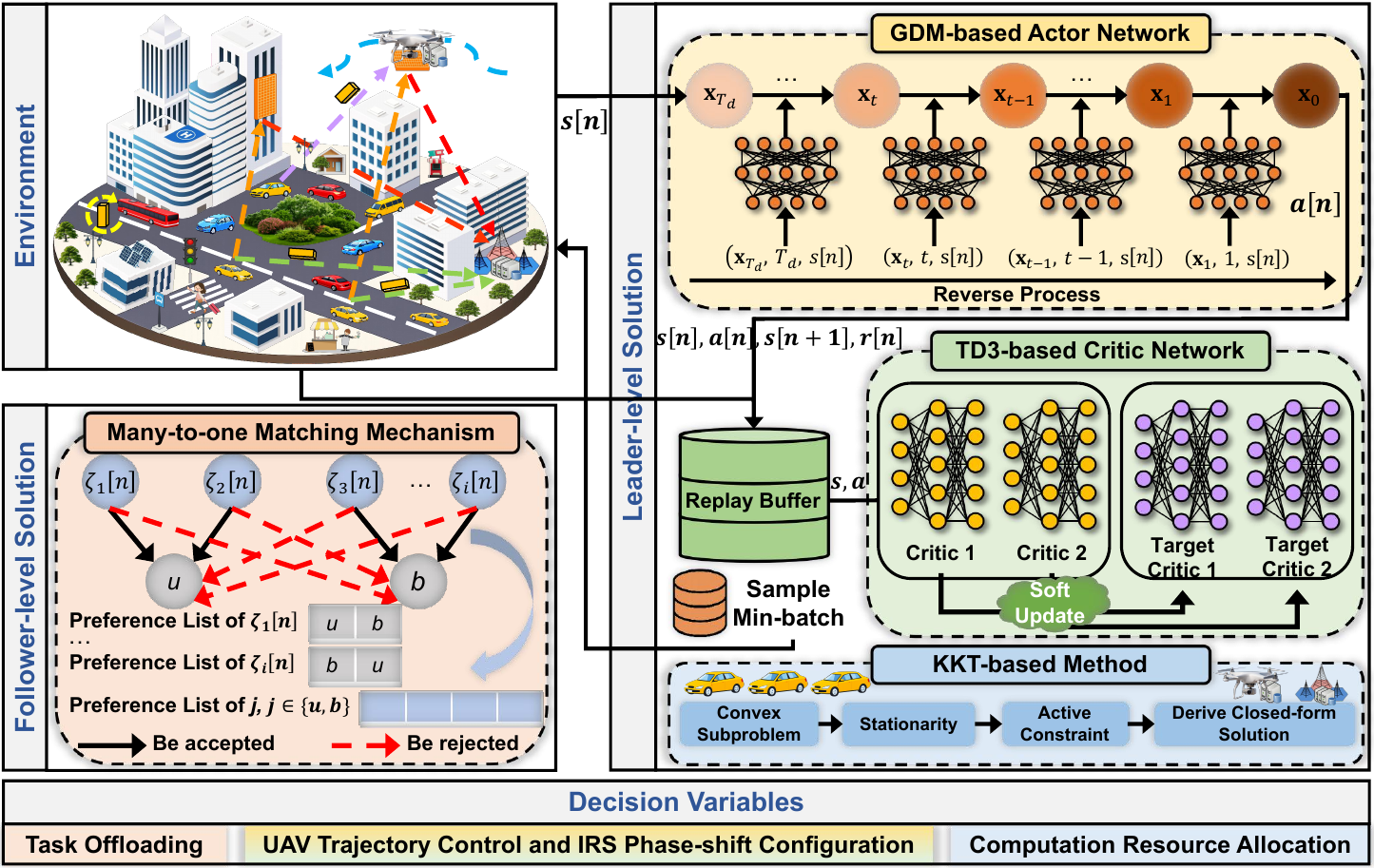}
    \vspace{-1.2em}
    \caption{The proposed HOOA. The MOOP is reformulated as a Stackelberg game with follower-level and leader-level problems. At the follower level, the many-to-one matching mechanism is employed to make task offloading decisions for vehicles. At the leader level, the GDMTD3 algorithm is leveraged to determine UAV trajectory control, IRS phase-shift configuration for MEC servers, while the KKT-based method is integrated to decide the computation resource allocation for MEC servers.}
    \label{fig_HOOA}
    \vspace{-2em}
\end{figure*}

 In this section, we first introduce the motivation for the proposed HOOA. Then, we reformulate MOOP as a Stackelberg game and present the corresponding follower-level and leader-level solutions. Furthermore, we summarize the main steps of HOOA and analyze its computational complexity. The framework of the proposed HOOA is shown in Fig.~\ref{fig_HOOA}, and the details are as follows.

\vspace{-1em}
\subsection{Motivation}
\label{sec:Motivation}

\par The motivations for developing HOOA are summarized as follows.
\begin{itemize}


   \item \textit{Reformulating the MOOP as a Stackelberg game.} To address the multi-objective trade-offs across heterogeneous decision entities, we observe that the IRS-enabled low-altitude MEC architecture inherently exhibits a hierarchical decision-making structure. Specifically, MEC servers determine system-level service-control decisions under a broader view of network states, while vehicles react by making individual decisions based on limited local information. Such an intrinsic hierarchy naturally gives rise to a leader-follower interaction between MEC servers and vehicles~\cite{Dong2024}. Therefore, we reformulate MOOP as a Stackelberg game to decompose it into follower-level and leader-level problems, thereby enabling structured coordination of conflicting objectives with improved scalability.

    
    \item \textit{Introducing a many-to-one matching mechanism for the follower-level discrete decision-making.} To tackle the NP-hard mixed-integer nature of MOOP, we focus on the discrete task offloading decisions at the follower level. Although each vehicle aims to minimize its individual cost, the offloading decisions are intrinsically coupled because each MEC server can serve only a limited number of concurrent tasks under the CPU-core constraint. Such coupling implies that optimizing task offloading independently could be infeasible at the system level. However, joint optimization of the task offloading decisions requires global information and iterative coordination, thus leading to high signaling overhead and decision latency. To overcome these challenges, we employ a many-to-one matching mechanism~\cite{Gu2015} that coordinates the associations between vehicles and MEC servers based on their preference relations. This mechanism yields feasible and scalable follower responses with low computational overhead.


   \item \textit{Enhancing the leader-level control with GDMTD3 algorithm and a KKT-based method.} To address the long-horizon and inter-slot coupling characteristics of MOOP, we design a learning-based solution for leader-level decision-making of MEC servers. Specifically, the control of UAV trajectories and IRS phase-shift configurations involves high-dimensional continuous actions with strong temporal dependence, which makes conventional short-term optimization ineffective. In this case, DRL-based methods are well suited for optimizing long-term performance through continuous interaction with dynamic environments~\cite{Li2022}. To further improve learning efficiency and stability, we incorporate GDM into TD3 to enhance action representation for UAV trajectory control and IRS configuration. Moreover, a KKT-based closed-form solution is derived for computation resource allocation, which reduces the effective action dimensionality and accelerates convergence.

\end{itemize}
\vspace{-0.8em}
\subsection{Problem Reformulation}
\label{sec:Problem_Reformulation}
\vspace{-0.2em}
\par In this subsection, we reformulate the MOOP as a Stackelberg game by specifying the cost functions of the vehicles and MEC servers, and then present the Stackelberg game problem.

\vspace{-0.5em}
\subsubsection{Vehicle Cost}
\par For each vehicle, we define a cost function that captures the task completion delay and energy consumption incurred at the vehicle side. Specifically, the delay-related cost of vehicle $i$ in time slot $n$ is defined as
\begin{equation}
\label{eq:cost_vehicle_T}
    C_i^{\mathrm{T}}(n) = \sum_{a\in\mathcal{A}} O_i^a(n)\, T_i^a(n).
\end{equation}

\par The corresponding energy-related cost of vehicle $i$ in time slot $n$ is defined as
\begin{equation}
\label{eq:cost_vehicle_E}
    C_i^{\mathrm{E}}(n) = O_i^i(n)\, E_i^i(n) + \sum_{j\in\{u,b\}} O_i^j(n)\, p_i^{\text{tr}}(n)D_i(n)/R_{i,j}(n).
\end{equation}

\par Accordingly, the overall cost of vehicle $i$ in time slot $n$ is expressed as
\begin{equation}
\label{eq:cost_vehicle}
    C_i(n) = \omega_i\, C_i^{\mathrm{T}}(n) + (1 - \omega_i)\, C_i^{\mathrm{E}}(n),
\end{equation}
\noindent where $\omega_i\in[0,1]$ is a weighting factor that controls the trade-off between delay cost and energy cost at the vehicle side.

\vspace{-1em}
\subsubsection{Server Cost}
\par Different from vehicles that are typically selfish, MEC servers are operated by a service provider that is responsible for the end-to-end quality-of-service (QoS) experienced by vehicles. Therefore, we define the cost of the MEC servers from a service-provisioning perspective, which includes not only the task completion delay and energy consumption incurred by MEC servers when processing offloaded tasks, but also the task completion delay and energy consumption incurred by vehicles during task offloading when receiving MEC services. Let $s$ denote the MEC servers as a whole. Specifically, the delay-related cost of the MEC servers in time slot $n$ is defined as
\begin{equation}
\label{eq:cost_server_T}
    C_{s}^{\mathrm{T}}(n) = \sum_{i \in \mathcal{I}} \sum_{j \in \{u,b\}} O_i^j(n)\, T_i^j(n).
\end{equation}

\par The corresponding energy-related cost of the MEC servers in time slot $n$ is defined as
\begin{equation}
\label{eq:cost_server_E}
    C_{s}^{\mathrm{E}}(n) = \sum_{i \in \mathcal{I}} \sum_{j \in \{u,b\}} O_i^j(n)\, E_i^{j}(n) + E_u^{\text{fly}}(n).
\end{equation}

\par Accordingly, the overall cost of the MEC servers in time slot $n$ can be expressed as
\begin{equation}
\label{eq:cost_server}
    C_{s}(n) = \omega_{s}\, C_{s}^{\mathrm{T}}(n) + (1 - \omega_{s})\, C_{s}^{\mathrm{E}}(n),
\end{equation}
\noindent where $\omega_{s}\in[0,1]$ is a weighting factor that controls the trade-off between delay cost and energy cost at the server side.

\vspace{-1em}
\subsubsection{Stackelberg Game Formulation}
\label{sec:stackelberg_game_formulation}
\par In a Stackelberg game, the leader \textcolor{b}{can make} decisions and announcing them before the followers, while the followers observe the decisions of the leader and respond with the optimal decisions~\cite{Wang2021}. In this case, \textcolor{b}{we reformulate the original MOOP} as a leader-follower Stackelberg game, where the MEC servers collectively act as the leader and the vehicles act as the followers. Based on this, MOOP is further divided into leader-level problem and follower-level problem, which are detailed as follows.

\par The follower-level problem for all vehicles in time slot $n$ is formulated as
\begin{subequations}
\label{eq:problem_follower_joint}
\begin{align}
    \mathbf{P}_f:\quad
    & \min_{\mathbf{O}} \ \sum_{i\in\mathcal{I}} C_i(n) \label{eq:problem_follower_joint_obj}\\
    & \text{s.t.}\quad \eqref{eq:constraint_offload_binary},\
    \eqref{eq:constraint_offload_unique},\
    \eqref{eq:constraint_deadline},\
    \eqref{eq:constraint_core_limit}. \label{eq:problem_follower_joint_cons}
\end{align}
\end{subequations}

\par Moreover, in Stackelberg game, each vehicle $i$ acts as an individual follower and aims to minimize its own cost in slot $n$, which can be expressed as
\begin{subequations}
\label{eq:problem_follower_ind}
\begin{align}
    \mathbf{P}_i:\quad
    & \min_{\mathbf{O}_i(n)} \ C_i(n) \label{eq:problem_follower_ind_obj}\\
    & \text{s.t.}\quad \eqref{eq:constraint_offload_binary},\
    \eqref{eq:constraint_offload_unique},\
    \eqref{eq:constraint_deadline}. \label{eq:problem_follower_ind_cons}
\end{align}
\end{subequations}

\par The leader-level problem for the MEC servers is formulated as
\begin{subequations}
\label{eq:problem_leader}
\begin{align}
    \mathbf{P}_l: \quad
    & \min_{\mathbf{Q}, \boldsymbol{\theta}, \mathbf{F}} \ C_s(n) \label{eq:problem_leader_obj} \\
    & \ \text{s.t.} \quad \eqref{eq:constraint_phase_bounds},\ \eqref{eq:constraint_cpu_sum},\ \eqref{eq:constraint_mobility_models} \label{eq:problem_leader_cons1}.
\end{align}
\end{subequations}

\vspace{-1em}
\subsection{Follower-Level Solution}
\label{sec:follower_solution}
\par In this subsection, we focus on the follower-level problem in the Stackelberg game by adopting a many-to-one matching mechanism to obtain the task offloading decision. Specifically, \textcolor{b}{the definitions and preliminaries of the matching model are presented.} Then, the preference lists are built for tasks and MEC servers. Based on these, matching construction presents an iterative request-and-admission procedure.

\vspace{-0.5em}
\subsubsection{\textcolor{b}{Definitions and Preliminaries}}
\label{sec:matching_fundamentals}

\par Let $\zeta^{\mathrm{req}}(n)=\{\zeta_i(n)\mid i\in\mathcal{I}\}$ denote the set of tasks generated by the vehicles in time slot $n$. Moreover, the task offloading decision for the tasks in $\zeta^{\mathrm{req}}(n)$ is determined by using a many-to-one matching mechanism, which is defined in Definition~\ref{def:matching_definition}.
\vspace{-0.5em}
\begin{definition}
\label{def:matching_definition}
In time slot $n$, the current matching is defined as a triplet $\big(\boldsymbol{\Omega}(n),\boldsymbol{\Phi}(n),\boldsymbol{\Upsilon}(n)\big)$.
\begin{itemize}
    \item $\boldsymbol{\Omega}(n)=\big(\zeta^{\mathrm{req}}(n),\{u,b\}\big)$ consists of the tasks and MEC servers.

    \item $\boldsymbol{\Phi}(n)=\big(\boldsymbol{\Phi}_{\zeta}(n),\boldsymbol{\Phi}_{j}(n)\big)$ consists of the preference lists of the tasks and MEC servers. Each task $\zeta_i(n)\in\zeta^{\mathrm{req}}(n)$ has a descending ordered preference list over the MEC servers, i.e., $\boldsymbol{\Phi}_{\zeta_i(n)}(n)=\{j\mid j\in\{u,b\},\, j\succ_{\zeta_i(n)} j'\}$, where $\succ_{\zeta_i(n)}$ denotes the preference of task $\zeta_i(n)$ towards the MEC servers. Moreover, each MEC server $j\in\{u,b\}$ has a descending ordered preference list over the tasks, i.e., $\boldsymbol{\Phi}_{j}(n)=\{\zeta_i(n)\in\zeta^{\mathrm{req}}(n),\, \zeta_i(n)\succ_{j}\zeta_{i'}(n)\}$.

    \item $\boldsymbol{\Upsilon}(n)\subseteq \zeta^{\mathrm{req}}(n)\times\{u,b\}$ is the matching between the tasks and MEC servers. Each task $\zeta_i(n)\in\zeta^{\mathrm{req}}(n)$ can be matched with at most one MEC server, i.e., $\boldsymbol{\Upsilon}_{\zeta_i(n)}(n)\in\{u,b\}$, while each MEC server $j\in\{u,b\}$ can be matched with multiple tasks, i.e., $\boldsymbol{\Upsilon}_{j}(n)\subseteq \zeta^{\mathrm{req}}(n)$.
\end{itemize}
\end{definition}

\vspace{-1em}
\subsubsection{Preference List Construction}
\label{sec:preference_list_construction}
\par In each time slot $n\in\mathcal{N}$, the preference lists in $\boldsymbol{\Phi}(n)$ are constructed by evaluating the preference values of tasks and MEC servers. In particular, the vehicle-side preference only accounts for the transmission delay and transmission energy. Thus, the preference value of each task $\zeta_i(n)\in \zeta^{\mathrm{req}}(n)$ on each MEC server $j\in\{u,b\}$ is defined as
\begin{equation}
\label{eq:pref_value_vehicle}
    \varrho_{\zeta_i(n)}^{j}(n)
    = -\big(\omega_i D_i(n)/R_{i,j}(n) + \big(1-\omega_i\big)p_i^{\text{tr}}(n)D_i(n)/R_{i,j}(n)\big).
\end{equation}
\noindent Note that a larger $\varrho_{\zeta_i(n)}^{j}(n)$ indicates a higher preference of task $\zeta_i(n)$ for MEC server $j$. Furthermore, the preference list $\boldsymbol{\Phi}_{\zeta_i(n)}(n)$ is constructed by ranking $\{\varrho_{\zeta_i(n)}^{j}(n)\}_{j\in\{u,b\}}$ in descending order.

\par The server-side preference only accounts for the computation energy. Thus, the preference value of each MEC server $j\in\{u,b\}$ on each task $\zeta_i(n)\in \zeta^{\mathrm{req}}(n)$ is defined as
\begin{equation}
\label{eq:pref_value_server}
    \varrho_{j}^{\zeta_i(n)}(n) = -\varpi_j D_i(n)G_i(n).
\end{equation}
\noindent Similarly, the preference list $\boldsymbol{\Phi}_{j}(n)$ is constructed by ranking $\{\varrho_{j}^{\zeta_i(n)}(n)\}_{\zeta_i(n)\in\zeta^{\mathrm{req}}(n)}$ in descending order.

\vspace{-0.2em}
\subsubsection{Matching Construction}
\label{sec:matching_construction}

\par Based on the preference lists in $\boldsymbol{\Phi}(n)$, the many-to-one matching $\boldsymbol{\Upsilon}(n)$ is constructed via an iterative request-and-admission procedure. The key steps are summarized below.
\begin{itemize}

    \item For each task $\zeta_i(n)\in\zeta^{\mathrm{rej}}(n)$ that is currently unmatched and has a nonempty preference list, the currently most preferred MEC server is selected as
    \begin{equation}
    \label{eq:match_select_server}
        j'=\boldsymbol{\Phi}_{\zeta_i(n)}(n)[1].
    \end{equation}

    \item Given the selected MEC server $j'$, the tentative matching for task $\zeta_i(n)$ is set as
    \begin{equation}
    \label{eq:match_tentative_task}
        \boldsymbol{\Upsilon}_{\zeta_i(n)}(n)=j'.
    \end{equation}
    Meanwhile, task $\zeta_i(n)$ is appended to the set of tasks currently associated with MEC server $j'$ as
    \begin{equation}
    \label{eq:match_tentative_server}
        \boldsymbol{\Upsilon}_{j'}(n)=\boldsymbol{\Upsilon}_{j'}(n)\cup\{\zeta_i(n)\}.
    \end{equation}

    \item Then, the tentative task-MEC server pair is added to the current matching set as
    \begin{equation}
    \label{eq:match_add_pair}
        \boldsymbol{\Upsilon}(n)=\boldsymbol{\Upsilon}(n)\cup\{(\zeta_i(n),j')\}.
    \end{equation}

    \item For each MEC server $j\in\{u,b\}$ that receives new requests, the set of rejected tasks is determined by retaining up to $m_{j}^{\mathrm{core}}$ most preferred tasks according to $\boldsymbol{\Phi}_{j}(n)$ as
    \begin{equation}
    \label{eq:match_reject_set}
        \mathcal{R}_{j}(n)=\boldsymbol{\Upsilon}_{j}(n)\setminus
    \operatorname{Top}_{m_{j}^{\mathrm{core}}}\!\big(\boldsymbol{\Upsilon}_{j}(n);\boldsymbol{\Phi}_{j}(n)\big).
    \end{equation}
    \textcolor{b}{\noindent where $\operatorname{Top}_{m_{j}^{\mathrm{core}}}(\cdot)$ denotes the set of up to $m_{j}^{\mathrm{core}}$ most preferred tasks in $\boldsymbol{\Upsilon}_{j}(n)$ according to $\boldsymbol{\Phi}_{j}(n)$.}

    \item The tentative acceptance set of MEC server $j$ is updated by removing the rejected tasks as
    \begin{equation}
    \label{eq:match_update_acceptance}
        \boldsymbol{\Upsilon}_{j}(n)=\boldsymbol{\Upsilon}_{j}(n)\setminus\mathcal{R}_{j}(n).
    \end{equation}

    \item The rejected tasks are added to $\zeta^{\mathrm{rej}}(n)$ for reconsideration in subsequent iterations as
    \begin{equation}
    \label{eq:match_update_rejset}
        \zeta^{\mathrm{rej}}(n)=\zeta^{\mathrm{rej}}(n)\cup\mathcal{R}_{j}(n).
    \end{equation}

    \item For each rejected task, MEC server $j$ is removed from the task preference list to avoid repeated requests to the same MEC server as
    \begin{equation}
    \label{eq:match_remove_server_from_pref}
        \boldsymbol{\Phi}_{\zeta_i(n)}(n)=\boldsymbol{\Phi}_{\zeta_i(n)}(n)\setminus\{j\}.
    \end{equation}

    \item Then, the tentative association is cleared for each rejected task and the corresponding task-MEC server pair is removed from the current matching set as
    \begin{equation}
    \label{eq:match_clear_task_match}
        \boldsymbol{\Upsilon}_{\zeta_i(n)}(n)=\emptyset.
    \end{equation}
    \begin{equation}
    \label{eq:match_remove_pair}
        \boldsymbol{\Upsilon}(n)=\boldsymbol{\Upsilon}(n)\setminus\{(\zeta_i(n),j)\}.
    \end{equation}

\end{itemize}

\par \textcolor{b}{The above updates are repeated until no unmatched task in $\zeta^{\mathrm{rej}}(n)$ has any MEC server left to request}, as shown in Algorithm~\ref{algo_matching}. After termination, $\boldsymbol{\Upsilon}(n)$ provides the resulting matching between tasks and MEC servers, and any task that remains unmatched is processed locally by default.

\begin{algorithm}[t]
\footnotesize
\SetAlgoLined

\textbf{Initialization:} $\zeta^{\mathrm{rej}}(n)\leftarrow\zeta^{\mathrm{req}}(n)$, $\boldsymbol{\Upsilon}(n)\leftarrow\emptyset$;

\For{$\zeta_i(n)\in\zeta^{\mathrm{req}}(n)$}{
    Compute and rank the preference values \textcolor{b}{by} Eq.~\eqref{eq:pref_value_vehicle} to obtain $\boldsymbol{\Phi}_{\zeta_i(n)}(n)$\;
}
\For{$j\in\{u,b\}$}{
    Compute and rank the preference values \textcolor{b}{by} Eq.~\eqref{eq:pref_value_server} to obtain $\boldsymbol{\Phi}_{j}(n)$\;
}

\While{There exists $\zeta_i(n)\in\zeta^{\mathrm{rej}}(n)$ such that $\boldsymbol{\Phi}_{\zeta_i(n)}(n)\neq\emptyset$ and $\boldsymbol{\Upsilon}_{\zeta_i(n)}(n)=\emptyset$}{
    \For{$\zeta_i(n)\in\zeta^{\mathrm{rej}}(n)$ such that $\boldsymbol{\Phi}_{\zeta_i(n)}(n)\neq\emptyset$ and $\boldsymbol{\Upsilon}_{\zeta_i(n)}(n)=\emptyset$}{
        Update the selected MEC server $j'$ \textcolor{b}{by} Eq.~\eqref{eq:match_select_server}\;
        Update $\boldsymbol{\Upsilon}_{\zeta_i(n)}(n)$ and $\boldsymbol{\Upsilon}_{j'}(n)$ \textcolor{b}{by} Eqs.~\eqref{eq:match_tentative_task}\textendash\eqref{eq:match_tentative_server}\;
        Update $\boldsymbol{\Upsilon}(n)$ \textcolor{b}{by} Eq.~\eqref{eq:match_add_pair}\;
    }
    \For{$j\in\{u,b\}$ \rm{that receives new requests}}{
        Update $\mathcal{R}_{j}(n)$, $\boldsymbol{\Upsilon}_{j}(n)$, and $\zeta^{\mathrm{rej}}(n)$ \textcolor{b}{by} Eqs.~\eqref{eq:match_reject_set}\textendash\eqref{eq:match_update_rejset}\;
        \For{$\zeta_i(n)\in\mathcal{R}_{j}(n)$}{
            Update $\boldsymbol{\Phi}_{\zeta_i(n)}(n)$ \textcolor{b}{by} Eq.~\eqref{eq:match_remove_server_from_pref}\;
            Update $\boldsymbol{\Upsilon}_{\zeta_i(n)}(n)$ and $\boldsymbol{\Upsilon}(n)$ \textcolor{b}{by} Eqs.~\eqref{eq:match_clear_task_match}\textendash\eqref{eq:match_remove_pair}\;
        }
    }
}
\caption{Matching for task offloading}
\label{algo_matching}
\end{algorithm}

\vspace{-1em}
\subsection{Leader-Level Solution}
\label{sec:leader_solution}
\par In this subsection, we first formulate the leader-level decision-making process in the Stackelberg game as a partially observable Markov decision process (POMDP). Subsequently, we develop a GDMTD3 algorithm to generate high-quality continuous actions for UAV trajectory control and IRS phase-shift configuration. Moreover, we introduce a KKT-based method to obtain the computation resource allocation efficiently.
\vspace{-0.6em}
\subsubsection{POMDP for the Stackelberg Game}
\label{sec:pomdp_stackelberg}
\par We characterize the leader-level decision-making process in the proposed Stackelberg game by using a POMDP framework~\cite{Zhang2023a} to capture the corresponding environment evolution across time slots. The POMDP is specified by the tuple $(\mathcal{S},\mathcal{A},\mathcal{P},\mathcal{R},\gamma)$, where $\mathcal{S}$, $\mathcal{A}$, $\mathcal{P}$, $\mathcal{R}$, and $\gamma$ denote the state space, action space, transition probability, reward function, and discount factor, respectively. At each time slot $n$, the environment is at state $\mathbf{s}(n)\in\mathcal{S}$, and the agent selects an action $\mathbf{a}(n)\in\mathcal{A}$ according to its policy. The environment then returns an instantaneous reward $r(n)=\mathcal{R}\big(\mathbf{s}(n),\mathbf{a}(n)\big)$ and transitions to the next state $\mathbf{s}(n+1)$ following the transition probability $\mathcal{P}\big(\mathbf{s}(n+1)\mid \mathbf{s}(n),\mathbf{a}(n)\big)$. Accordingly, the key elements of the POMDP are described below.

\par \textbf{1) State Space.} The environment state is constructed as a concatenated vector that integrates the current system information with the interaction history, which can be defined as $\mathbf{s}(n)=\big\{\allowbreak \mathbf{q}_u(n),\allowbreak \mathbf{q}_i(n),\allowbreak D_i(n),\allowbreak G_i(n),\allowbreak T_i^{\max}(n),\allowbreak \mathbf{O}^{\mathrm{his}}(n),\allowbreak \mathbf{Q}^{\mathrm{his}}(n),\allowbreak \boldsymbol{\theta}^{\mathrm{his}}(n),\allowbreak \mathbf{F}^{\mathrm{his}}(n)\ \big|\ \forall i\in\mathcal{I},\ j\in\{u,b\}\big\}$. Here, $\mathbf{q}_u(n)$ denotes the horizontal position of the UAV, $\mathbf{q}_i(n)$ represents the horizontal position of vehicle $i$, and $\langle D_i(n),\, G_i(n),\, T_i^{\max}(n) \rangle$ specifies the task attributes of vehicle $i$. Moreover, $\mathbf{O}^{\mathrm{his}}(n)$, $\mathbf{Q}^{\mathrm{his}}(n)$, $\boldsymbol{\theta}^{\mathrm{his}}(n)$, and $\mathbf{F}^{\mathrm{his}}(n)$ collect the most recent $\varsigma$-slot histories of the follower decisions and the leader decisions, respectively. Note that in the initial environment, these history records can be randomly generated.

\par \textbf{2) Action Space.} The action space corresponds to the set of decisions of the leader, i.e., the MEC servers. Therefore, the action space is defined as $\mathbf{a}(n)=\big\{\allowbreak v_u(n),\allowbreak \varphi_u(n),\allowbreak \theta_{k,l}(n),\allowbreak f_{j,i}(n)\ \big|\ \forall i\in\mathcal{I},\ j\in\{u,b\},\ k\in\mathcal{K},\ l\in\mathcal{L}\big\}.$

\par \textbf{3) Reward Function.} The reward evaluates the effectiveness of the leader action in terms of the overall cost of the considered system. Specifically, the instantaneous reward in time slot $n$ is defined as the negative weighted sum of the total vehicle cost and the server cost, together with the penalty for constraint violations, which is given as
\begin{equation}
\label{eq:reward_def}
    r(n) = -\big(\omega_{c}\,\sum_{i\in \mathcal{I}}{C}_{i}(n) + \big(1-\omega_{c}\big)\,{C}_{s}(n)\big) - r^{\mathrm{bv}}(n) - r^{\mathrm{dv}}(n),
\end{equation}
\noindent where $\omega_{c}\in[0,1]$ is a weighting factor that balances the total vehicle cost and server cost in the reward, $r^{\mathrm{bv}}(n)$ denotes the penalty for UAV boundary violations, and $r^{\mathrm{dv}}(n)$ indicates the penalty for task deadline violations, which is given as
\begin{equation}
\label{eq:penalty_dv}
    r^{\mathrm{dv}}(n) = \sum_{i\in\mathcal{I}} \nu_i(n)\, r_i(n),
\end{equation}
\noindent where $\nu_i(n)$ is a binary variable that represents whether task $\zeta_i(n)$ violates its deadline in slot $n$, and $r_i(n)$ is the corresponding penalty.

\par \textbf{4) POMDP Analysis.} In the above POMDP framework, the leader action consists of UAV trajectory control, IRS phase-shift configuration, and computation resource allocation. Despite the strong capability of DRL in sequential decision-making, learning a unified policy over these coupled continuous variables remains challenging due to the high action dimensionality and stringent coupling constraints, which may lead to slow convergence and unstable training~\cite{Luong2019}. Consequently, we apply GDMTD3 algorithm to improve learning efficiency and stability by jointly optimizing the UAV trajectory control and IRS phase-shift configuration, as described in Subsection~\ref{sec:GDM_TD3_Algorithm}. Moreover, we adopt a KKT-based method to reduce the effective action dimensionality by analytically deriving the computation resource allocation, as detailed in Subsection~\ref{sec:KKT_based_Method}.
\vspace{-0.8em}
\subsubsection{GDMTD3 Algorithm}
\label{sec:GDM_TD3_Algorithm}
\par \textbf{1) Standard TD3 Algorithm.} TD3 algorithm is an actor-critic DRL method for continuous control that improves deep deterministic policy gradient (DDPG) by mitigating Q-value overestimation and enhancing training stability~\cite{Chen2023}. We present the principles of the standard TD3 algorithm as follows.

\par \textit{\textbf{Actor-Critic Structure.}} TD3 algorithm employs an actor-critic framework, where neural networks are used to approximate a deterministic policy and the corresponding Q-functions. Specifically, the deterministic policy is represented by an actor network $\mu(\mathbf{s}\,|\,\boldsymbol{\eta})$, and the value evaluation is performed by a double-Q critic consisting of two independent critic networks $Q_{1}(\mathbf{s},\mathbf{a}\,|\,\boldsymbol{\zeta}_{1})$ and $Q_{2}(\mathbf{s},\mathbf{a}\,|\,\boldsymbol{\zeta}_{2})$, where $\boldsymbol{\eta}$, $\boldsymbol{\zeta}_{1}$, and $\boldsymbol{\zeta}_{2}$ denote the parameters of the actor and two critic networks, respectively. To enhance training stability, TD3 algorithm further maintains target networks, namely $\mu'(\mathbf{s}\,|\,\boldsymbol{\eta}')$, $Q_{1}'(\mathbf{s},\mathbf{a}\,|\,\boldsymbol{\zeta}_{1}')$, and $Q_{2}'(\mathbf{s},\mathbf{a}\,|\,\boldsymbol{\zeta}_{2}')$, where $\boldsymbol{\eta}'$, $\boldsymbol{\zeta}_{1}'$, and $\boldsymbol{\zeta}_{2}'$ denote the parameters of the target networks.

\par \textit{\textbf{Network Update.}} The critic networks are updated to minimize the discrepancy between the current Q-values and target Q-value. Thus, the loss function of the critic networks is given as
\begin{equation}
\label{eq:critic_loss}L(\boldsymbol{\zeta}_1,\boldsymbol{\zeta}_2)=\mathbb{E}\![(Q_1(\mathbf{s},\mathbf{a}\,|\,\boldsymbol{\zeta}_1)-y^{\mathrm{tv}})^2+(Q_2(\mathbf{s},\mathbf{a}\,|\,\boldsymbol{\zeta}_2)-y^{\mathrm{tv}})^2],
\end{equation}
\noindent where $y^{\mathrm{tv}} = r(n) + (1-d^{\mathrm{done}}(n))\gamma \min\{Q_1'(\mathbf{s}(n+1),\tilde{\mathbf{a}}(n+1)\,|\,\boldsymbol{\zeta}_1'),\,Q_2'(\mathbf{s}(n+1),\tilde{\mathbf{a}}(n+1)\,|\,\boldsymbol{\zeta}_2')\}$.

\par Following the delayed policy update strategy in the TD3 algorithm, the actor network is updated by minimizing
\begin{equation}
\label{eq:actor_loss}
    L(\boldsymbol{\eta})=-\mathbb{E}\!\left[\,Q_1\!\big(\mathbf{s},\mu(\mathbf{s}\,|\,\boldsymbol{\eta})\,\big|\,\boldsymbol{\zeta}_1\big)\right].
\end{equation}

\par After each delayed actor update, the target networks are softly updated as
\begin{subequations}
\label{eq:soft_update}
\begin{align}
    \boldsymbol{\zeta}_1' &\leftarrow \tau\,\boldsymbol{\zeta}_1 + \big(1-\tau\big)\,\boldsymbol{\zeta}_1', \label{eq:soft_update_a}\\
    \boldsymbol{\zeta}_2' &\leftarrow \tau\,\boldsymbol{\zeta}_2 + \big(1-\tau\big)\,\boldsymbol{\zeta}_2', \label{eq:soft_update_b}\\
    \boldsymbol{\eta}' &\leftarrow \tau\,\boldsymbol{\eta} + \big(1-\tau\big)\,\boldsymbol{\eta}', \label{eq:soft_update_c}
\end{align}
\end{subequations}
\noindent where $\tau$ controls the soft-update rate.

\par \textbf{2) GDM-based Actor Network.} To effectively cope with the complexity and uncertainty of decision generation in the considered Stackelberg game, we employ the denoising diffusion probabilistic model (DDPM)~\cite{ho2020denoising} to construct a GDM-based actor network. Specifically, the DDPM consists of a forward noising process and a reverse denoising process. This iterative denoising mechanism enables deep modeling of the underlying decision distribution and supports the generation of increasingly refined continuous decisions under the current state. The mathematical representation of the DDPM is given as follows.

\par \textit{\textbf{Forward Process.}} For a given original data $\mathbf{x}_0$, the forward process generates a sequence of noisy samples $\{\mathbf{x}_t\}_{t=1}^{T_{\mathrm{d}}}$ by progressively injecting Gaussian noise. The transition from $\mathbf{x}_{t-1}$ to $\mathbf{x}_t$ is governed by the conditional distribution $q(\mathbf{x}_t\,|\,\mathbf{x}_{t-1})$, which is given as
\begin{equation}
\label{eq:ddpm_forward_transition}
    q\big(\mathbf{x}_t\,|\,\mathbf{x}_{t-1}\big)=\mathcal{N}\!\big(\mathbf{x}_t;\sqrt{1-\beta_t}\,\mathbf{x}_{t-1},\,\beta_t\mathbf{I}\big),
\end{equation}
\noindent where $\beta_t = 1 - e^{-\beta^{\min}/T_{\mathrm{d}} - (2t-1)/(2T_{\mathrm{d}}^2)(\beta^{\max}-\beta^{\min})}$ is the variational posterior schedule, and $\mathbf{I}$ represents the identity matrix.

\par Thus, the forward process from $\mathbf{x}_0$ to $\mathbf{x}_{T_{\mathrm{d}}}$ is written as
\begin{equation}
\label{eq:ddpm_forward_chain}
    q\big(\mathbf{x}_{1:T_{\mathrm{d}}}\,|\,\mathbf{x}_0\big)=\prod_{t=1}^{T_{\mathrm{d}}} q\big(\mathbf{x}_t\,|\,\mathbf{x}_{t-1}\big).
\end{equation}

\par However, as the value of $t$ increases, obtaining $\mathbf{x}_t$ by repeatedly applying Eq.~\eqref{eq:ddpm_forward_transition} incurs a computational overhead that scales linearly with $t$. To avoid this sequential sampling, we exploit the Gaussian structure of the forward transitions and directly express $\mathbf{x}_t$ in terms of $\mathbf{x}_0$ as
\begin{equation}
\label{eq:ddpm_forward_closed_form}
    \mathbf{x}_t=\sqrt{\bar{\alpha}_t}\,\mathbf{x}_0+\sqrt{1-\bar{\alpha}_t}\,\boldsymbol{\epsilon},
\end{equation}
\noindent where $\boldsymbol{\epsilon}\sim\mathcal{N}(\mathbf{0},\mathbf{I})$, $\alpha_t = 1-\beta_t$, and $\bar{\alpha}_t = \prod_{\ell=1}^{t}\alpha_{\ell}$ denotes the cumulative product of $\{\alpha_{\ell}\}_{\ell=1}^{t}$.

\par Note that the abovementioned forward process is defined on the original data $\mathbf{x}_0$, which is an optimal solution to the optimization problem. However, in the considered IRS-enabled low-altitude MEC architecture for vehicular networks, such an optimal $\mathbf{x}_0$ is generally unavailable in advance. Hence, we mainly leverage the subsequent reverse process to construct the actor network.

\par \textit{\textbf{Reverse Process.}} The reverse process removes the injected noise from $\mathbf{x}_{T_{\mathrm{d}}}$ and recovers the original data $\mathbf{x}_0$ via a sequence of Gaussian transitions. However, evaluating $q(\mathbf{x}_{t-1}\mid \mathbf{x}_t)$ requires the real data distribution, which is intractable in practice. To address this issue, we employ a parameterized model $p_{\boldsymbol{\delta}}$ to approximate $q(\mathbf{x}_{t-1}\mid \mathbf{x}_t)$, which is expressed as
\begin{equation}
\label{eq:ddpm_reverse_transition}
    p_{\boldsymbol{\delta}}\big(\mathbf{x}_{t-1}\,|\,\mathbf{x}_t\big)=\mathcal{N}\!\big(\mathbf{x}_{t-1};\,\kappa_{\boldsymbol{\delta}}\big(\mathbf{x}_t,t,\mathbf{g}\big),\,\tilde{\beta}_t\mathbf{I}\big),
\end{equation}
\noindent where $\kappa_{\boldsymbol{\delta}}(\mathbf{x}_t,t,\mathbf{g}) =
\sqrt{\alpha_t}(1-\bar{\alpha}_{t-1})/(1-\bar{\alpha}_t)\,\mathbf{x}_t
+\sqrt{\bar{\alpha}_{t-1}}\beta_t/(1-\bar{\alpha}_t)\,\mathbf{x}_0$
and $\tilde{\beta}_t=(1-\bar{\alpha}_{t-1})\beta_t/(1-\bar{\alpha}_t)$ are the mean and variance for the denoising model, respectively.

\par However, the parameterized model $p_{\boldsymbol{\delta}}$ has no access to $\mathbf{x}_0$ and therefore relies on an estimate $\hat{\mathbf{x}}_0$, which is given as
\begin{equation}
\label{eq:ddpm_x0_est}
    \hat{\mathbf{x}}_0=1/\sqrt{\bar{\alpha}_t}\,\big(\mathbf{x}_t-\sqrt{1-\bar{\alpha}_t}\,\boldsymbol{\epsilon}_{\boldsymbol{\delta}}\big(\mathbf{x}_t,t,\mathbf{g}\big)\big),
\end{equation}
\noindent where $\boldsymbol{\epsilon}_{\boldsymbol{\delta}}(\mathbf{x}_t,t,\mathbf{g})$ denotes a deep neural network, and then $\kappa_{\boldsymbol{\delta}}(\mathbf{x}_t,t,\mathbf{g})$ can be expressed as
\begin{equation}
\label{eq:ddpm_reverse_mean_eps}
    \kappa_{\boldsymbol{\delta}}(\mathbf{x}_t,t,\mathbf{g})
    =1/\sqrt{\alpha_t}\,(\mathbf{x}_t-\beta_t/\sqrt{1-\bar{\alpha}_t}\,
    \boldsymbol{\epsilon}_{\boldsymbol{\delta}}(\mathbf{x}_t,t,\mathbf{g})).
\end{equation}
\par Thus, the reverse process from $\mathbf{x}_{T_{\mathrm{d}}}$ to $\mathbf{x}_0$ is written as
\begin{equation}
\label{eq:ddpm_reverse_chain}
    p_{\boldsymbol{\delta}}\big(\mathbf{x}_{0:T_{\mathrm{d}}}\big)=p\big(\mathbf{x}_{T_{\mathrm{d}}}\big)\prod_{t=1}^{T_{\mathrm{d}}} p_{\boldsymbol{\delta}}\big(\mathbf{x}_{t-1}\,|\,\mathbf{x}_t\big),
\end{equation}
\noindent where $p(\mathbf{x}_{T_{\mathrm{d}}})$ is a Gaussian distribution.

\par Consequently, the reverse process of the DDPM can be embedded into the actor network of the proposed GDMTD3 algorithm to generate high-quality continuous actions. Moreover, to facilitate gradient-based optimization of the actor network, the Gaussian transition in Eq.~\eqref{eq:ddpm_reverse_transition} is implemented in a reparameterized form, which is given as
\begin{equation}
\label{eq:ddpm_reparameterization}
    \mathbf{x}_{t-1}=\kappa_{\boldsymbol{\delta}}\big(\mathbf{x}_t,t,\mathbf{g}\big)+\mathbb{I}\big(t>0\big)\sqrt{\tilde{\beta}_t}\,\boldsymbol{\xi},
\end{equation}
\noindent where $\boldsymbol{\xi}\sim\mathcal{N}(\mathbf{0},\mathbf{I})$, and $\mathbb{I}(t>0)$ ensures that no noise is added at the final denoising step.

\vspace{-0.8em}
\subsubsection{KKT-based Method}
\label{sec:KKT_based_Method}
\par Given the decisions of task offloading $\bar{\mathbf{O}}$, UAV trajectory control $\bar{\mathbf{Q}}$ and IRS phase-shift configuration $\bar{\boldsymbol{\theta}}$, while removing the irrelevant constant terms, the leader-level problem is transformed into a computation resource allocation problem, which is expressed as
\begin{subequations}
\label{eq:problem_c}
\begin{align}
    \mathbf{P}_{l}^{\mathrm{c}}:\quad
    & \min_{\mathbf{F}} \
    \sum_{j\in\{u,b\}}\sum_{i\in\mathcal{I}}
    O_i^{j}(n)\; D_i(n)G_i(n)/f_{j,i}(n)
    \label{eq:problem_c_obj}\\
    \text{s.t.}\quad
    & \sum_{i\in\mathcal{I}} O_i^{j}(n)\; f_{j,i}(n) \le F_j^{\max}, \forall j \in \{u,b\}, \ n \in \mathcal{N}.
    \label{eq:problem_c_cons}
\end{align}
\end{subequations}

\par Problem $\mathbf{P}_{l}^{\mathrm{c}}$ is a convex optimization problem, as proved in Theorem~\ref{thm:plc_convex}. Therefore, standard convex optimization techniques can be employed to solve it. Moreover, by exploiting the KKT conditions, we further derive a closed-form expression of the optimal computation resource allocation, which is summarized in Theorem~\ref{thm:plc_optimal}.

\begin{theorem}
\label{thm:plc_convex}
Problem $\mathbf{P}_{l}^{\mathrm{c}}$ is a convex optimization problem.
\end{theorem}
\begin{myproof}
Refer to Section I of the supplementary material.
\end{myproof}

\begin{theorem}
\label{thm:plc_optimal}
The optimal computation resource allocation for problem $\mathbf{P}_{l}^{\mathrm{c}}$ is given as $\mathbf{F}^{\star}=\{f_{j,i}^{\star}(n)\mid \forall j\in\{u,b\},\, i\in\mathcal{I},\, n\in\mathcal{N}\}$, where
\begin{equation}
\label{eq:plc_optimal_f}
    f_{j,i}^{\star}(n)=\frac{\sqrt{D_i(n)G_i(n)/F_j^{\max}}\;F_j^{\max}}{\sum_{i\in\mathcal{I}}O_{i}^{j}(n)\sqrt{D_i(n)G_i(n)/F_j^{\max}}}.
\end{equation}
\end{theorem}

\begin{myproof}
Refer to Section II of the supplementary material.
\end{myproof}
\vspace{-1em}
\subsection{Main Steps of HOOA and Analysis}
\label{sec:HOOA_steps_analysis}
\par In this subsection, we introduce the main steps of the proposed HOOA approach and analyze its computational complexity.
\vspace{-1em}
\subsubsection{Main Steps of HOOA Approach}
\label{sec:HOOA_main_steps}
\par The proposed HOOA approach is outlined in Algorithm~\ref{algo_HOOA}. Specifically, HOOA solves the considered Stackelberg game by coordinating leader decisions of UAV trajectory control, IRS phase-shift configuration and computation resource allocation with follower decision of task offloading. During the training phase, the agent interacts with the environment over multiple episodes and time slots. At each time slot, the agent selects an action based on the current state under the policy parameterized by the GDM-based actor network. Then, the environment returns a reward and the next observation after executing the selected action together with the deterministic many-to-one matching mechanism and KKT-based method. For network updates, mini-batches are randomly sampled from the buffer to update the critic networks, while the actor network is updated in a delayed manner and the target networks are soft-updated to ensure training stability.

\begin{algorithm}[t]
\footnotesize
\SetAlgoLined
\SetKwRepeat{Repeat}{repeat}{until}

Initialize an experience replay buffer $\mathcal{D}$ and set the delayed update interval $d$\;
Initialize the online GDM-based actor network $\mu(\mathbf{s}\,|\,\boldsymbol{\eta})$, the critic networks $Q_1(\mathbf{s},\mathbf{a}\,|\,\boldsymbol{\zeta}_1)$ and $Q_2(\mathbf{s},\mathbf{a}\,|\,\boldsymbol{\zeta}_2)$, and the target networks $\mu'(\mathbf{s}\,|\,\boldsymbol{\eta}')$, $Q_1'(\mathbf{s},\mathbf{a}\,|\,\boldsymbol{\zeta}_1')$, and $Q_2'(\mathbf{s},\mathbf{a}\,|\,\boldsymbol{\zeta}_2')$\;

\For{each episode}{
    Reset environment, obtain the initial state $\mathbf{s}(0)$\;
    $step \leftarrow 0$\;
    $delay\_counter \leftarrow 0$\;

    \Repeat{environment is terminated or $step \ge N$}{
        Receive the current state $\mathbf{s}(step)$\;

        Sample $\mathbf{x}_{T_{\mathrm{d}}}\sim\mathcal{N}(\mathbf{0},\mathbf{I})$\;
        \For{each reverse denoising step $t=T_{\mathrm{d}},T_{\mathrm{d}}-1,\ldots,1$}{
            Obtain $\kappa_{\boldsymbol{\delta}}\big(\mathbf{x}_t,t,\mathbf{s}(step)\big)$ \textcolor{b}{by} Eq.~\eqref{eq:ddpm_reverse_mean_eps}\;
            Update $\mathbf{x}_{t-1}$ \textcolor{b}{by} Eq.~\eqref{eq:ddpm_reparameterization}\;
        }
        Set $\mathbf{a}(step)\leftarrow \mathbf{x}_0$\;
        Execute $\mathbf{a}(step)$ to update $\mathbf{Q}(step)$ and $\boldsymbol{\theta}(step)$\;

        Obtain $\mathbf{O}^{\star}(step)$ for task offloading by calling \textbf{Algorithm~\ref{algo_matching}}\;
        Compute $\mathbf{F}^{\star}(step)$ \textcolor{b}{by} Eq.~\eqref{eq:plc_optimal_f}\;

        Execute the joint decision $\big(\mathbf{a}(step),\mathbf{O}^{\star}(step),\mathbf{F}^{\star}(step)\big)$\;
        Observe the reward $r(step)$ and the next state $\mathbf{s}(step+1)$\;

        Store $\big(\mathbf{s}(step),\mathbf{a}(step),r(step),\mathbf{s}(step+1)\big)$ into $\mathcal{D}$\;

        Randomly sample a mini-batch $\mathcal{B}$ from $\mathcal{D}$\;
        Update the critic networks \textcolor{b}{by} Eq.~\eqref{eq:critic_loss}\;

        $delay\_counter \leftarrow delay\_counter + 1$\;
        \If{$delay\_counter > d$}{
            Update the actor network \textcolor{b}{by} Eq.~\eqref{eq:actor_loss}\;
            Soft-update the target networks \textcolor{b}{by} Eq.~\eqref{eq:soft_update}\;
            $delay\_counter \leftarrow 0$\;
        }

        $step \leftarrow step + 1$\;
    }
}
\caption{The Proposed HOOA}
\label{algo_HOOA}
\end{algorithm}
\vspace{-1.5em}
\subsubsection{Complexity Analysis}
\label{sec:HOOA_complexity}
\vspace{-0.5em}
\par The complexity of the proposed HOOA approach is examined for the training phase and execution phase.

\par \textit{\textbf{Training Phase.}} During the training phase, the computational complexity at each time slot is primarily determined by experience collection and network updates. For experience collection, action selection under the policy parameterized by the GDM-based actor network requires $T_{\mathrm{d}}$ reverse denoising steps, thereby resulting in a computation complexity of $\mathcal{O}\big(T_{\mathrm{d}}|\boldsymbol{\eta}|\big)$\textcolor{b}{\cite{Zhang2025}}. Meanwhile, executing the deterministic many-to-one matching mechanism for task offloading incurs $\mathcal{O}\big(|\mathcal{I}|\log|\mathcal{I}|\big)$, and the KKT-based method for computation resource allocation has a linear complexity $\mathcal{O}\big(|\mathcal{I}|\big)$\textcolor{b}{\cite{Ning2022}}. Hence, the complexity of experience collection is $\mathcal{O}\big(T_{\mathrm{d}}|\boldsymbol{\eta}|+|\mathcal{I}|\log|\mathcal{I}|+|\mathcal{I}|\big)$. For network updates, each critic update with a mini-batch of size $\mathcal{B}$ includes generating target actions by the target GDM-based actor network with $T_{\mathrm{d}}$ denoising steps and updating the two critic networks, which leads to $\mathcal{O}\big(\mathcal{B}T_{\mathrm{d}}|\boldsymbol{\eta}|+\mathcal{B}\big(|\boldsymbol{\zeta}_1|+|\boldsymbol{\zeta}_2|\big)\big)$\textcolor{b}{\cite{Zhang2025}}. Moreover, the actor network is updated once every $d+1$ steps, and thus its average computational complexity per step is $\mathcal{O}\big(\big(\mathcal{B}T_{\mathrm{d}}|\boldsymbol{\eta}|+\mathcal{B}\big(|\boldsymbol{\zeta}_1|+|\boldsymbol{\zeta}_2|\big)\big)/(d+1)\big).$ Additionally, the  space complexity is mainly attributed to the network parameters and the replay buffer, i.e., $\mathcal{O}\big(|\boldsymbol{\eta}|+|\boldsymbol{\zeta}_1|+|\boldsymbol{\zeta}_2|+|\mathcal{D}|\big(|\mathbf{s}|+|\mathbf{a}|\big)\big)$.

\par \textit{\textbf{Execution Phase.}} During the execution phase, no network update is performed and the computational complexity at each time slot mainly comes from action selection and deterministic decision computation\textcolor{b}{\cite{Fujimoto2018}}, which is $\mathcal{O}\big(T_{\mathrm{d}}|\boldsymbol{\eta}|+|\mathcal{I}|\log|\mathcal{I}|+|\mathcal{I}|\big)$, while the space complexity is $\mathcal{O}\big(|\boldsymbol{\eta}|\big)$ for storing the actor-network parameters.

\vspace{-1.2em}
\section{Simulation Results}
\label{sec:simulation_results}
\subsection{Simulation Setup}
\label{sec:simulation_setup}
\par In this section, simulation results are presented to validate the effectiveness of the proposed approach.

\par \textit{\textbf{Scenarios.}} We consider an IRS-enabled low-altitude MEC architecture for vehicular networks, where a UAV and a BS are deployed to provide offloading services to $10$ vehicles in a $1000\times 1000~\text{m}^2$ square area. To enhance the wireless link quality, we deploy two IRSs, i.e., a building-installed IRS and a UAV-carried IRS. Additionally, the system timeline is set to $T=100~\mathrm{s}$ and discretized into $N=100$ equal time slots. The default simulation parameters are summarized in Table~\ref{tab_simuParameter}.
\begin{table}[t]
\setlength{\abovecaptionskip}{-3pt}%
\centering
\begin{minipage}{0.48\textwidth} 
\centering
\caption{Simulation parameters}
\label{tab_simuParameter}
\renewcommand*{\arraystretch}{1.05}
\scriptsize
\setlength{\tabcolsep}{3pt}
\begin{tabular}{p{0.1\columnwidth}|p{0.52\columnwidth}|p{0.3\columnwidth}}
\hline
\hline
\multicolumn{3}{c}{\textbf{System Parameters}} \\
\hline
\textbf{Symbol} & \textbf{Description} & \textbf{Default value} \\
\hline
$H$ & Fixed altitude of the UAV & $100~\mathrm{m}$ \\
\hline
$\mathbf{q}_b$ & Position of the BS & $[800,200]~\mathrm{m}$ \\
\hline
$\mathbf{q}^{\mathrm{fix}}$ & Position of the building-installed IRS & $[200,800,75]~\mathrm{m}$ \\
\hline
$p_i^{\mathrm{tr}}$ & Transmit power of vehicle $i$ & $25~\mathrm{dBm}$ \\
\hline
$\sigma^2$ & Noise power & $-98~\mathrm{dBm}$\\
\hline
$\rho$ & Path loss at the reference distance $1$ m & $10^{-3}$ \\
\hline
$\gamma^{\mathrm{rf}}$ & Rician factor & $3~\mathrm{dB}$ \\
\hline
$L$ & Reflecting elements of each IRS & $64$ \\
\hline
$F_i^{\max}$ & Maximum computing capability of vehicle $i$ & $1~\mathrm{GHz}$ \\
\hline
$B_{i,j}$ & Bandwidth allocated to vehicle $i$ for MEC server $j$ & $10~\mathrm{MHz}\ (j=u),\ 20~\mathrm{MHz}\ (j=b)$ \\
\hline
$F_j^{\max}$ & Maximum computing capability of MEC server $j$ & $[10,20]~\mathrm{GHz}\ (j=u),\ [20,40]~\mathrm{GHz}\ (j=b)$ \\
\hline
$m_j^{\mathrm{core}}$ & CPU core number of MEC server $j$ & $[2,8]$ \\
\hline
$\kappa_i$ & Effective switched capacitance coefficient of the CPU in vehicle $i$ & $10^{-28}$ \\
\hline
$\varpi_j$ & Effective computation energy coefficient of MEC server $j$ & $8.2\times10^{-9}$ \\
\hline
$v_u^{\max}$ & Maximum speed of the UAV & $25~\mathrm{m/s}$ \\
\hline
$D_i$ & Task size & $[1,5]~\mathrm{Mb}$ \\
\hline
$G_i$ & Computation intensity of the task & $[500,1000]~\mathrm{cycles/bit}$ \\
\hline
$T_i^{\max}$ & Deadline of the task & $[1,5]~\mathrm{s}$ \\
\hline
\hline
\multicolumn{3}{c}{\textbf{Learning Parameters}} \\
\hline
\textbf{Symbol} & \textbf{Description} & \textbf{Default value} \\
\hline
$d$ & Delayed policy update interval in TD3 & $2$ \\
\hline
$\gamma$ & Discount factor & $0.99$ \\
\hline
$w$ & Hidden layer width & $400$ \\
\hline
$\alpha_{\mu},\ \alpha_{Q}$ & Learning rates of the actor and critic networks & $3\times10^{-4},\ 3\times10^{-4}$ \\
\hline
$\tau$ & Soft-update rate of the target networks & $5\times10^{-3}$ \\
\hline
$\mathcal{B}$ & Mini-batch size & $256$ \\
\hline
$T_{\mathrm{d}}$ & Number of diffusion timesteps in GDM & $10$ \\
\hline
\hline
\end{tabular}
\end{minipage}
\end{table}


\par \textit{\textbf{Benchmarks.}} The proposed HOOA is evaluated by comparing it with several other approaches and algorithms.

\begin{itemize}
    \item \textit{Nearby task offloading (NAO)}: the tasks of each vehicle are offloaded to the nearby MEC server in each time slot, while the remaining decision variables follow the proposed HOOA.

    \item \textit{Equal computation resource allocation (ECRA)}: the computation resources of each MEC server are equally allocated among the offloaded tasks in each time slot, while the remaining decision variables follow the proposed HOOA.

    \item \textit{Fixed IRS phase-shift configuration (FIPSC)}: the IRS phase-shift configuration is kept fixed throughout the whole horizon, while the remaining decision variables follow the proposed HOOA.

    \item \textit{Circular UAV trajectory control (CUTC)}: the UAV flies along a predefined circular trajectory, while the remaining decision variables follow the proposed HOOA.

    \item \textit{TD3}: the leader-level solution is learned by the TD3 algorithm, while the follower-level solution follows the proposed HOOA.
    
    \item \textit{DDPG}: the leader-level solution is learned by the DDPG algorithm, while the follower-level solution follows the proposed HOOA.
    
    \item \textit{Soft actor-critic (SAC)}: the leader-level solution is learned by the SAC algorithm, while the follower-level solution follows the proposed HOOA.

\end{itemize}

\vspace{-1em}
\subsection{Evaluation Results}
\label{sec:evaluation_results}

\subsubsection{System Performance Comparison with Different Approaches}
\label{sec:comparison_different_approaches}

\par
\begin{figure*}[!hbt]
    \centering
    \subfigure[]{
        \includegraphics[width=0.23\linewidth]{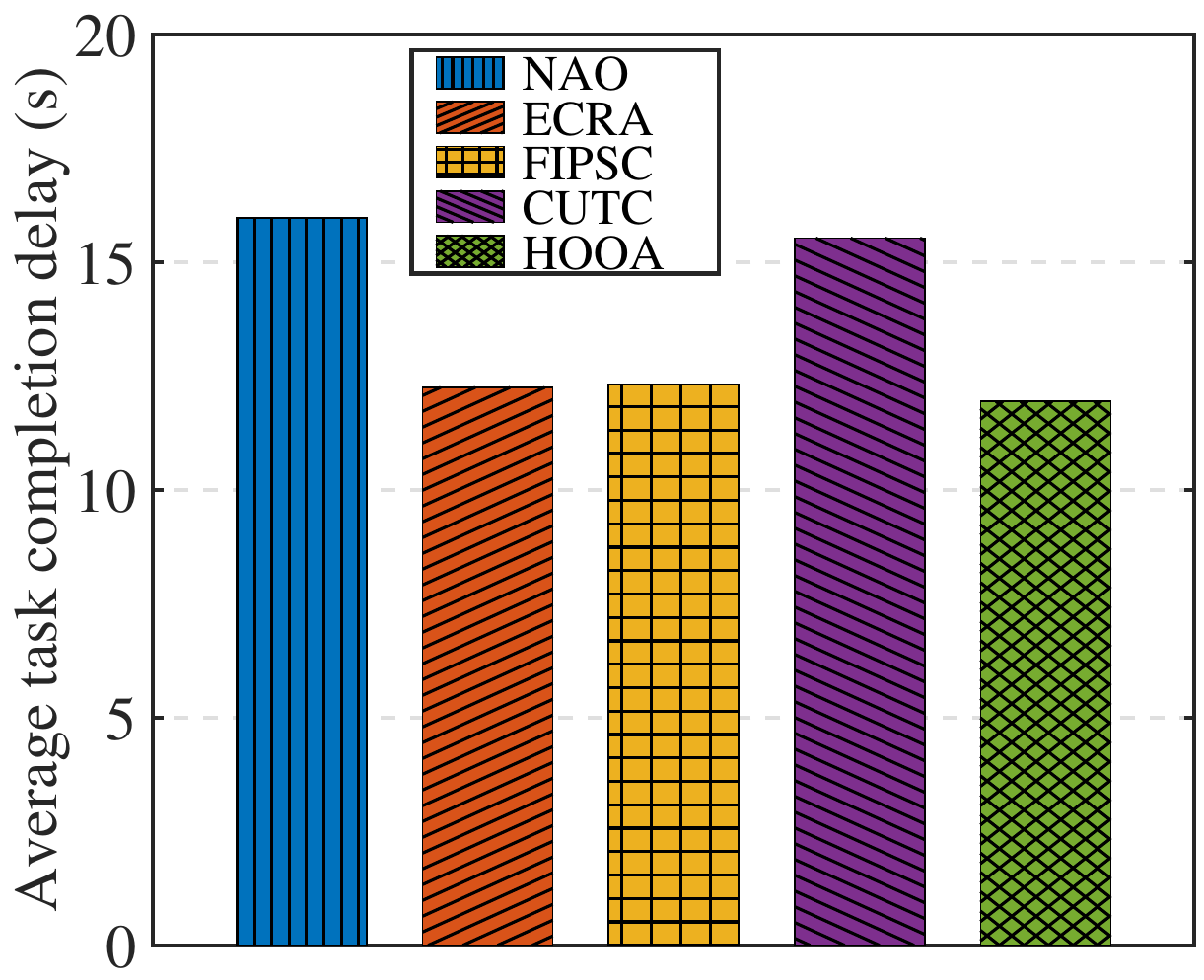}
    }\hfill
    \subfigure[]{
        \includegraphics[width=0.23\linewidth]{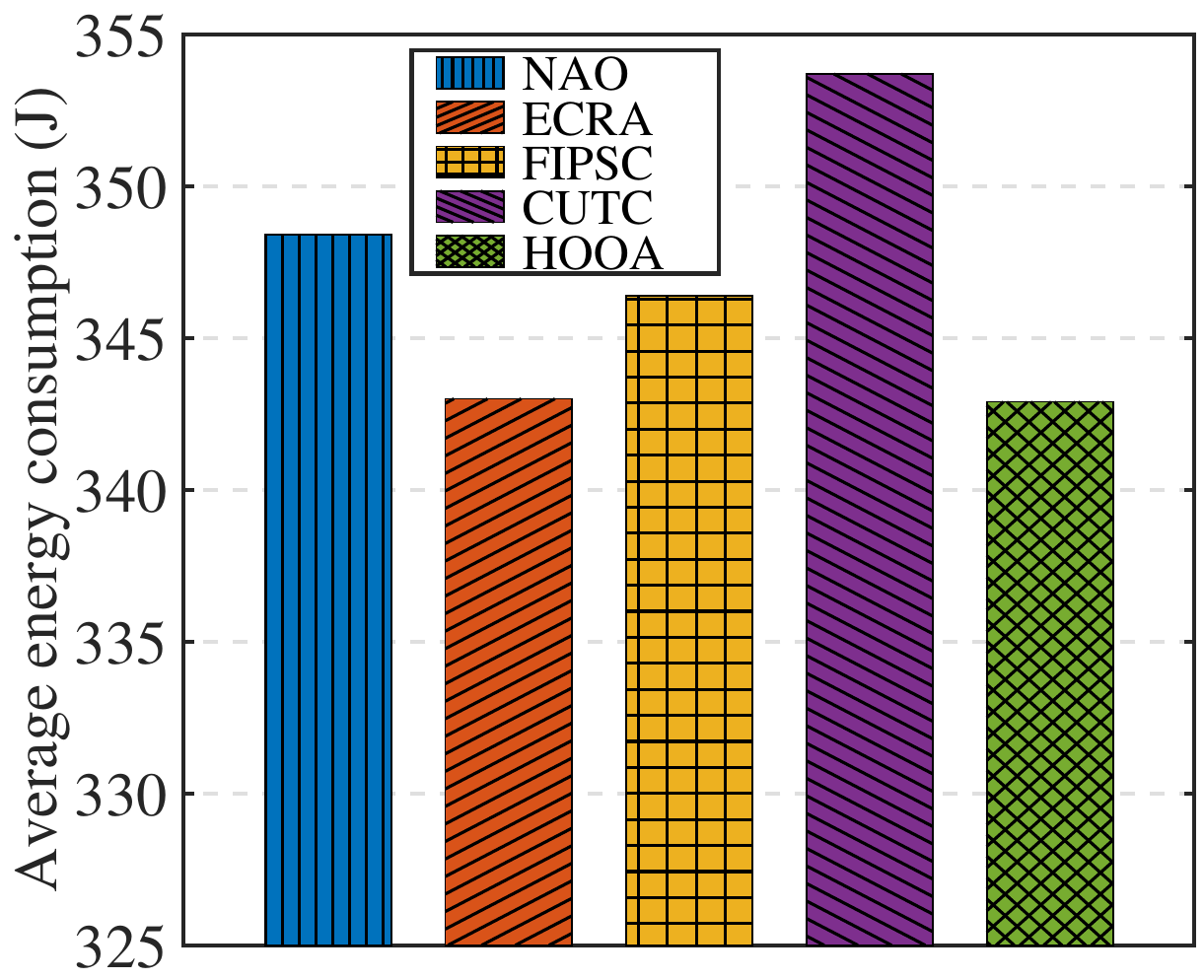}
    }\hfill
    \subfigure[]{
        \includegraphics[width=0.23\linewidth]{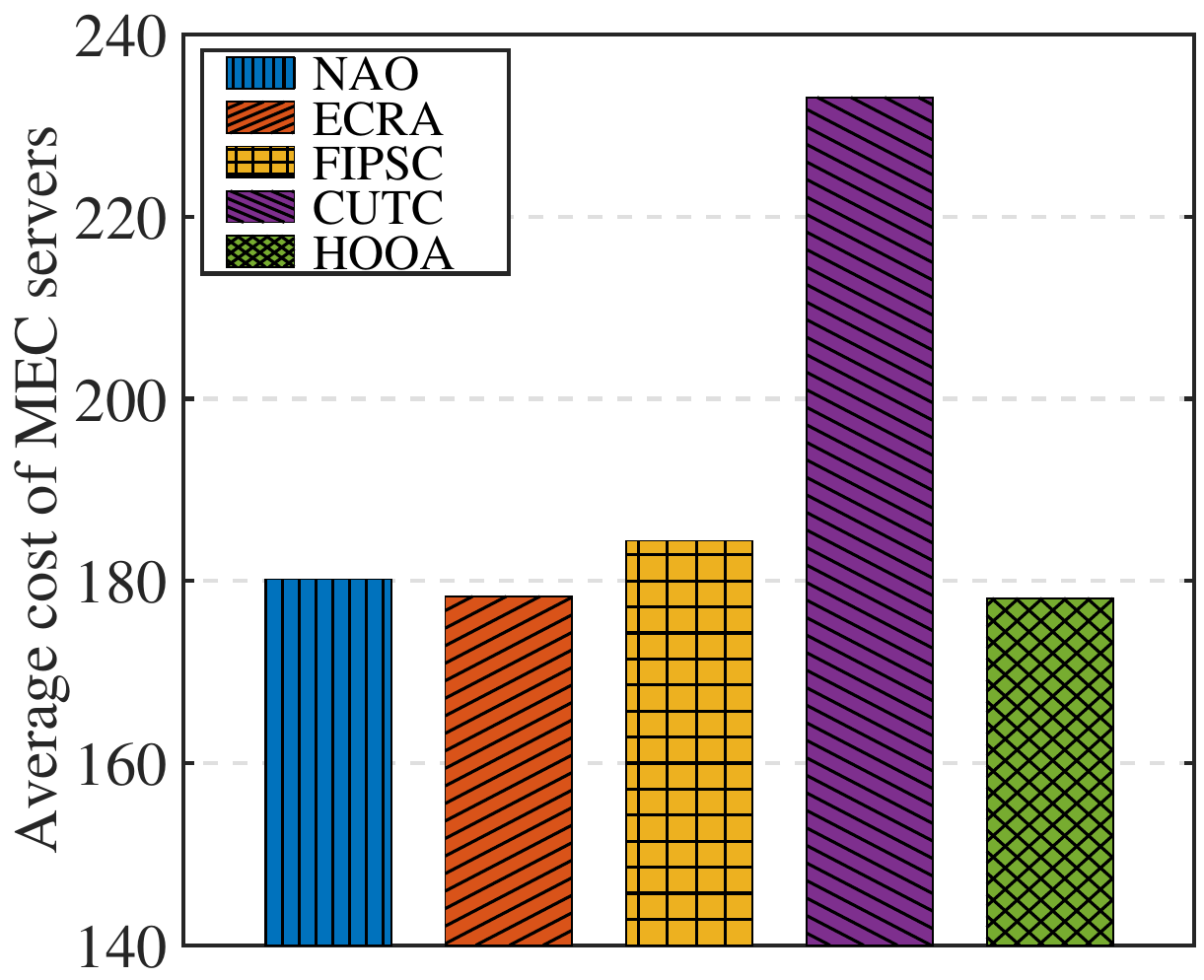}
    }\hfill
    \subfigure[]{
        \includegraphics[width=0.23\linewidth]{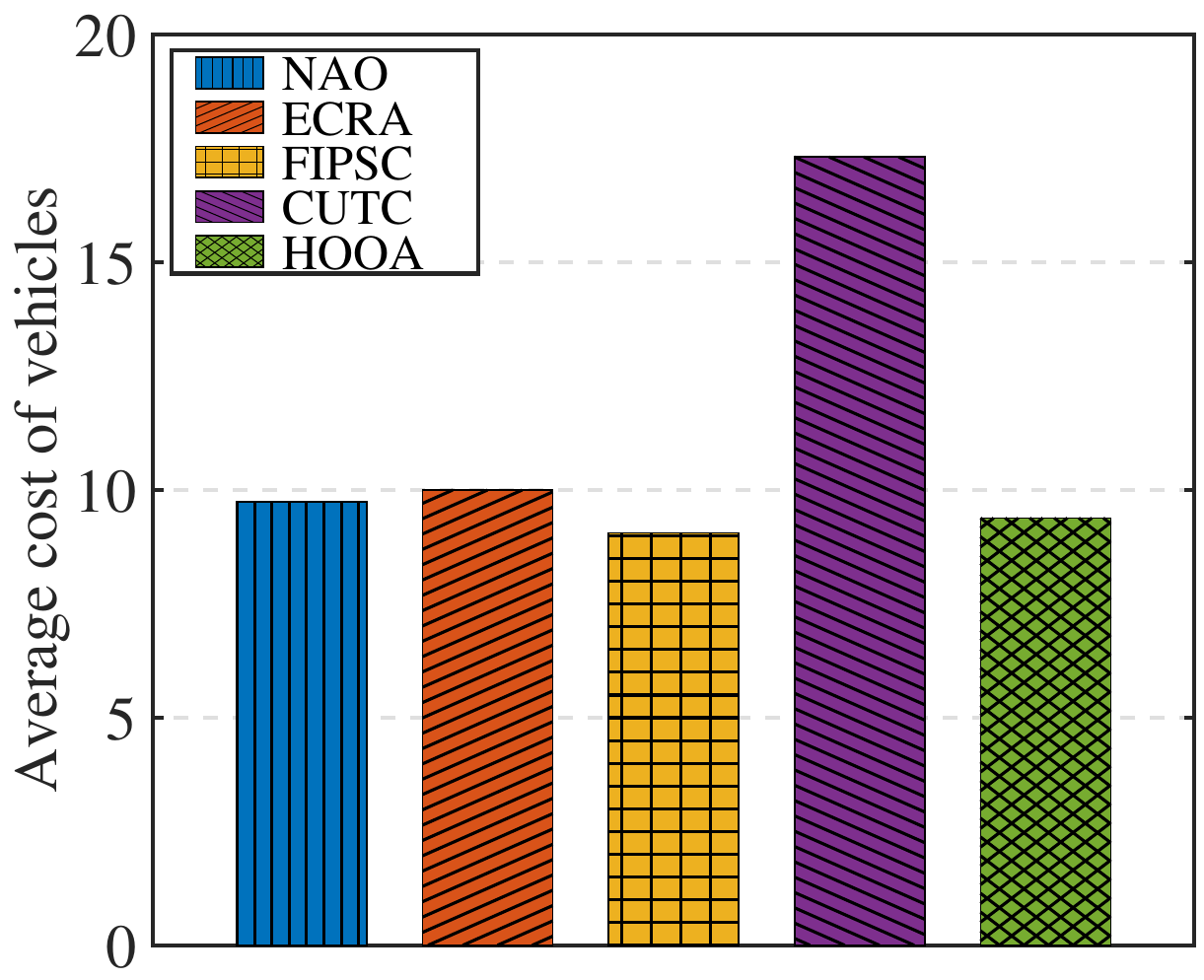}
    }
    \vspace{-1.2em}
    \caption{System performance comparison among different approaches. (a) Average task completion delay. (b) Average energy consumption. (c) Average cost of MEC servers. (d) Average cost of vehicles.}
    \label{fig_System_Performance}
    \vspace{-1.8em}
\end{figure*}

 Fig.~\ref{fig_System_Performance} compares the system performance of different approaches under the default parameter settings. As shown in Figs.~\ref{fig_System_Performance}(a), \ref{fig_System_Performance}(b), and \ref{fig_System_Performance}(c), the proposed HOOA approach significantly outperforms the benchmarks in terms of average task completion delay, average energy consumption, and average cost of MEC servers. These results indicate that jointly optimizing task offloading, UAV trajectory control, IRS phase-shift configuration, and computation resource allocation can effectively enhance the overall system efficiency in the considered IRS-enabled low-altitude MEC architecture for vehicular networks. In contrast, FIPSC and CUTC lack adaptability to time-varying air-ground channels and blockage conditions, thereby weakening the effectiveness of link-quality enhancement and leading to degraded delay and energy performance. Meanwhile, NAO and ECRA cannot efficiently utilize the limited computation resources of MEC servers in dynamic environments, which exacerbates the challenge of satisfying stringent vehicular quality of service requirements. 

\par Additionally, it can be observed from Figs.~\ref{fig_System_Performance}(b) and \ref{fig_System_Performance}(c) that ECRA achieves average energy consumption and average cost of MEC servers close to those of the proposed HOOA. This stems from the fact that the overall energy consumption is largely determined by the uplink transmission energy consumption and the UAV flight energy consumption, whereas equal CPU allocation mainly changes the task completion delay. Moreover, Fig.~\ref{fig_System_Performance}(d) shows that the proposed HOOA incurs a slightly higher average cost of vehicles than FIPSC. This is because HOOA adopts more proactive offloading to optimize overall system performance, thus increasing vehicle-side transmission expenditure. Nevertheless, this modest increase is accompanied by substantial reduction in the average task completion delay, the average energy consumption, and the average cost of MEC servers, thereby demonstrating a practical trade-off between vehicle-side cost and global service efficiency.

\par In summary, the results in Fig.~\ref{fig_System_Performance} highlight the necessity of jointly optimizing heterogeneous decision variables. Meanwhile, the proposed HOOA achieves significant reductions in average task completion delay, average energy consumption, and average cost of MEC servers.

\vspace{-0.5em}
\subsubsection{Convergence Comparison with State-of-the-Art DRL Algorithms}
\label{sec:comparison_sota_drl}
\par
\begin{figure*}[!hbt]
    \centering
    \subfigure[]{
        \includegraphics[width=0.3\linewidth]{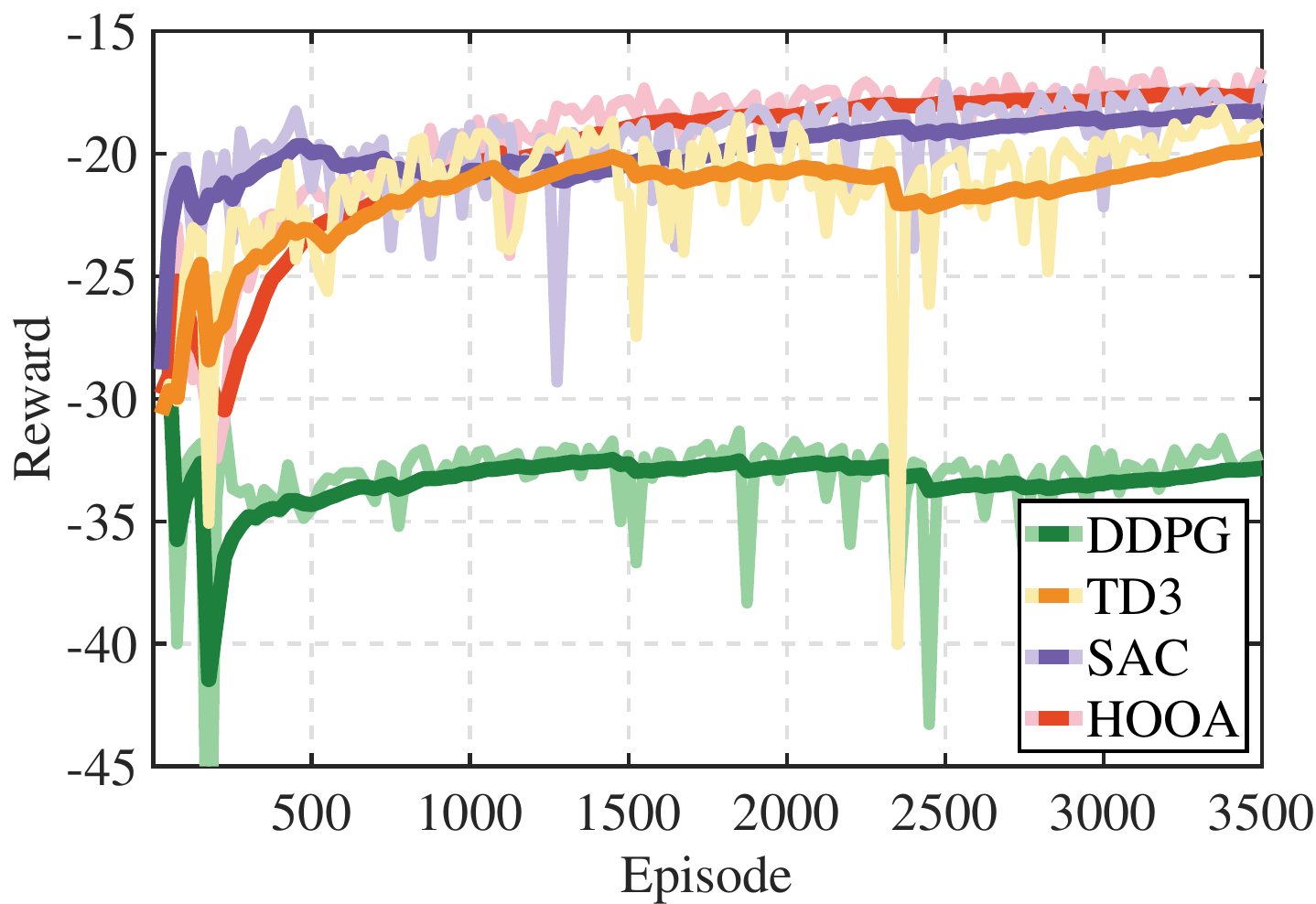}
    }\hfill
    \subfigure[]{
        \includegraphics[width=0.3\linewidth]{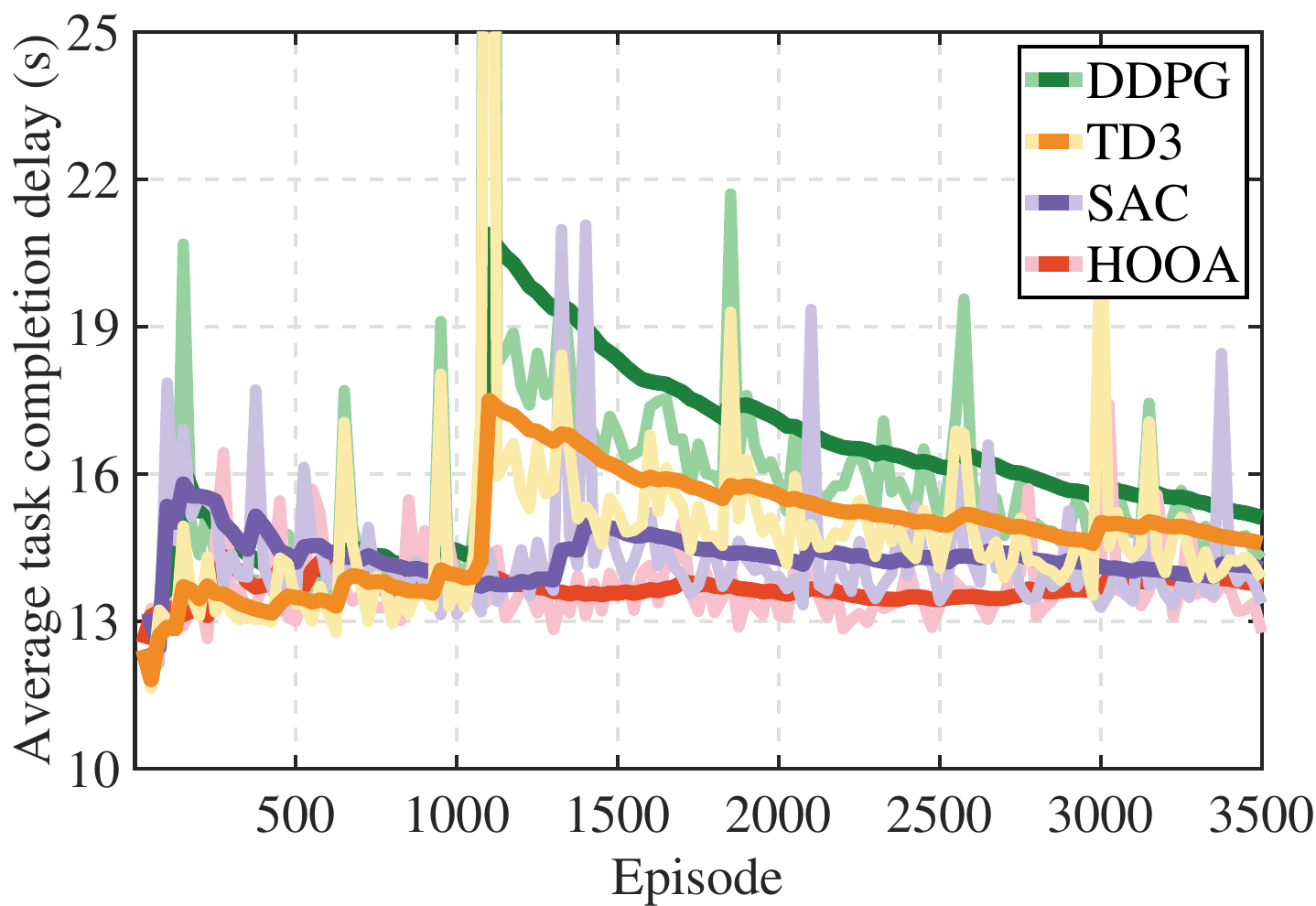}
    }\hfill
    \subfigure[]{
        \includegraphics[width=0.3\linewidth]{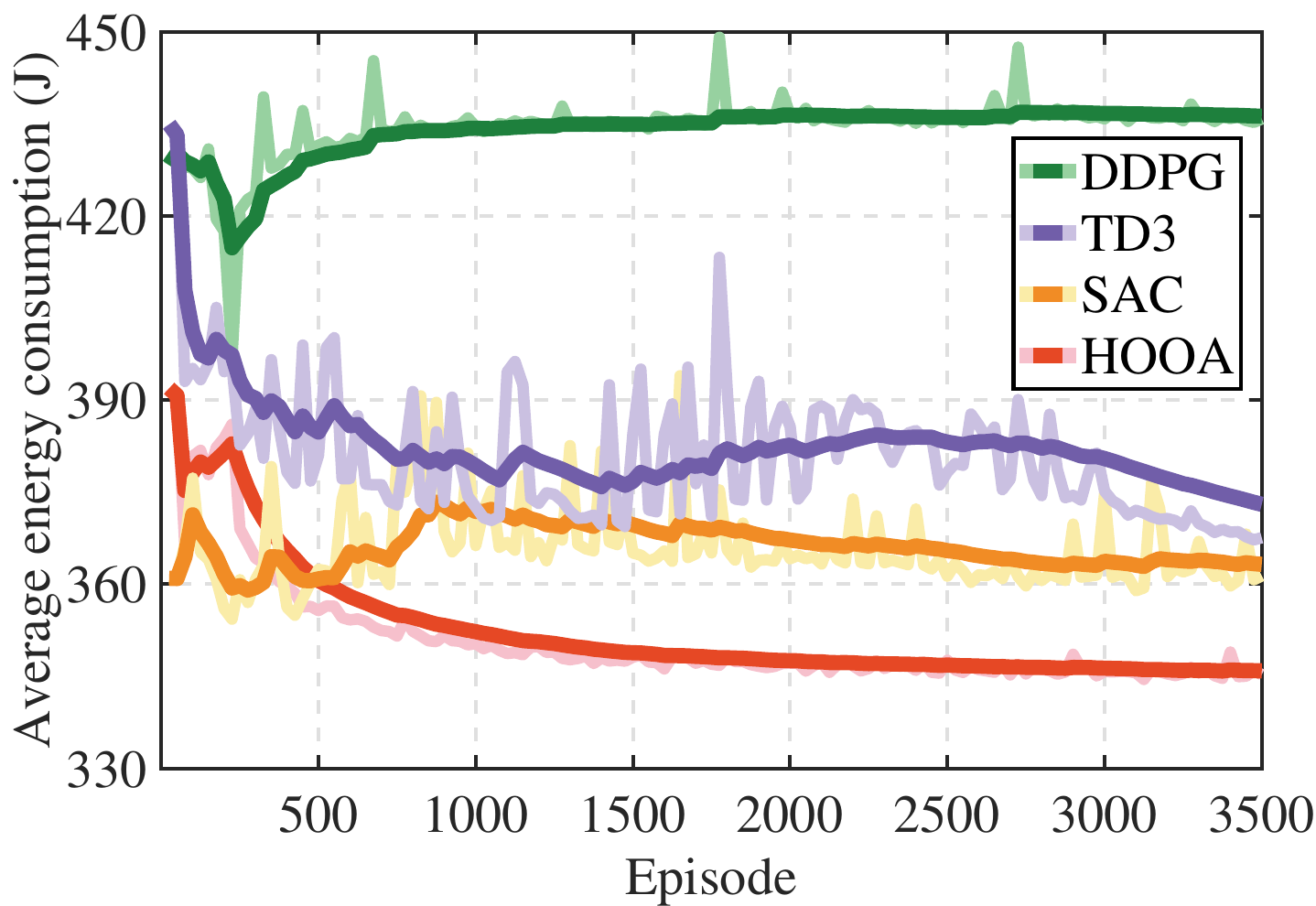}
    }
    \vspace{-1.2em}
\caption{Comparison of convergence curves of the proposed HOOA approach and some state-of-the-art DRL algorithms. (a) Reward. (b) Average task completion delay. (c) Average energy consumption.}
    \label{fig_Convergence}
    \vspace{-1.8em}
\end{figure*}

 Fig.~\ref{fig_Convergence}(a) presents the reward of the proposed HOOA in comparison with other state-of-the-art DRL algorithms during the training process. As can be seen, the proposed HOOA reaches the highest reward after convergence and exhibits the best stability among all the other algorithms. This performance advantage is primarily due to embedding the GDM-based iterative denoising generation mechanism into TD3, which produces diverse and high-quality continuous actions of joint UAV trajectory control and IRS phase-shift configuration, thereby improving exploration efficiency and policy optimization in time-varying air-ground environments. In contrast, DDPG attains the lowest reward and exhibits early convergence because its single-critic value estimate is highly error-sensitive, which often leads to earlier convergence to suboptimal policies. Moreover, TD3 improves over DDPG by mitigating Q-value overestimation, while still converges to a relatively low reward, since noise-based exploration is inefficient in a strongly coupled and high-dimensional continuous action space. In addition, although SAC achieves noticeable reward improvements in the early training stage and exhibits a smoother increase, it remains inferior to the proposed HOOA. This is mainly because SAC often depend on entropy-driven exploration, which struggles to consistently produce feasible actions under stringent system constraints.

\par Fig.~\ref{fig_Convergence}(b) depicts the average task completion delay per episode during the training process. It can be observed that the proposed HOOA converges faster and reaches the lowest and most stable delay. This indicates that the diffusion-based actor network supports steadier optimization and learns more reliable continuous control actions, which leads to more consistent delay performance. However, DDPG and TD3 converge to higher delay levels with more \textcolor{b}{significant} oscillations due to inefficient exploration in the strongly coupled continuous decision space. Moreover, SAC shows a smoother delay reduction and achieves competitive performance, but still remains slightly inferior to the proposed HOOA. A key reason is that entropy-driven exploration is less effective at producing stable and precise actions when fine-grained control is required.

\par Fig.~\ref{fig_Convergence}(c) shows the average energy consumption per episode during training. As can be observed, the energy consumption under the proposed HOOA continues to decrease throughout training and eventually converges to the lowest level with only minor fluctuations. In comparison, DDPG exhibits a distinctive pattern where the energy consumption drops briefly in the early stage and then rises to the highest plateau, which indicates that the learned policy gradually shifts toward more energy-intensive operating modes as training proceeds. Moreover, TD3 and SAC achieve substantially lower energy consumption than DDPG, but are still inferior to the proposed HOOA in terms of the converged energy level and the fluctuation magnitude. These results reveal that the proposed HOOA can better exploit the representation and exploration capabilities of the GDM to produce more stable continuous decisions, which encourages smoother UAV movement and more favorable link conditions, thereby reducing UAV flight energy consumption and transmission energy consumption.

\par Accordingly, it can be concluded that the proposed HOOA achieves the best overall convergence performance among the considered state-of-the-art DRL algorithms, as evidenced by the highest converged reward together with the lowest task completion delay and energy consumption.

\begin{figure*}[!hbt]
    \centering
    \subfigure[Diffusion step]{
        \includegraphics[width=0.3\linewidth]{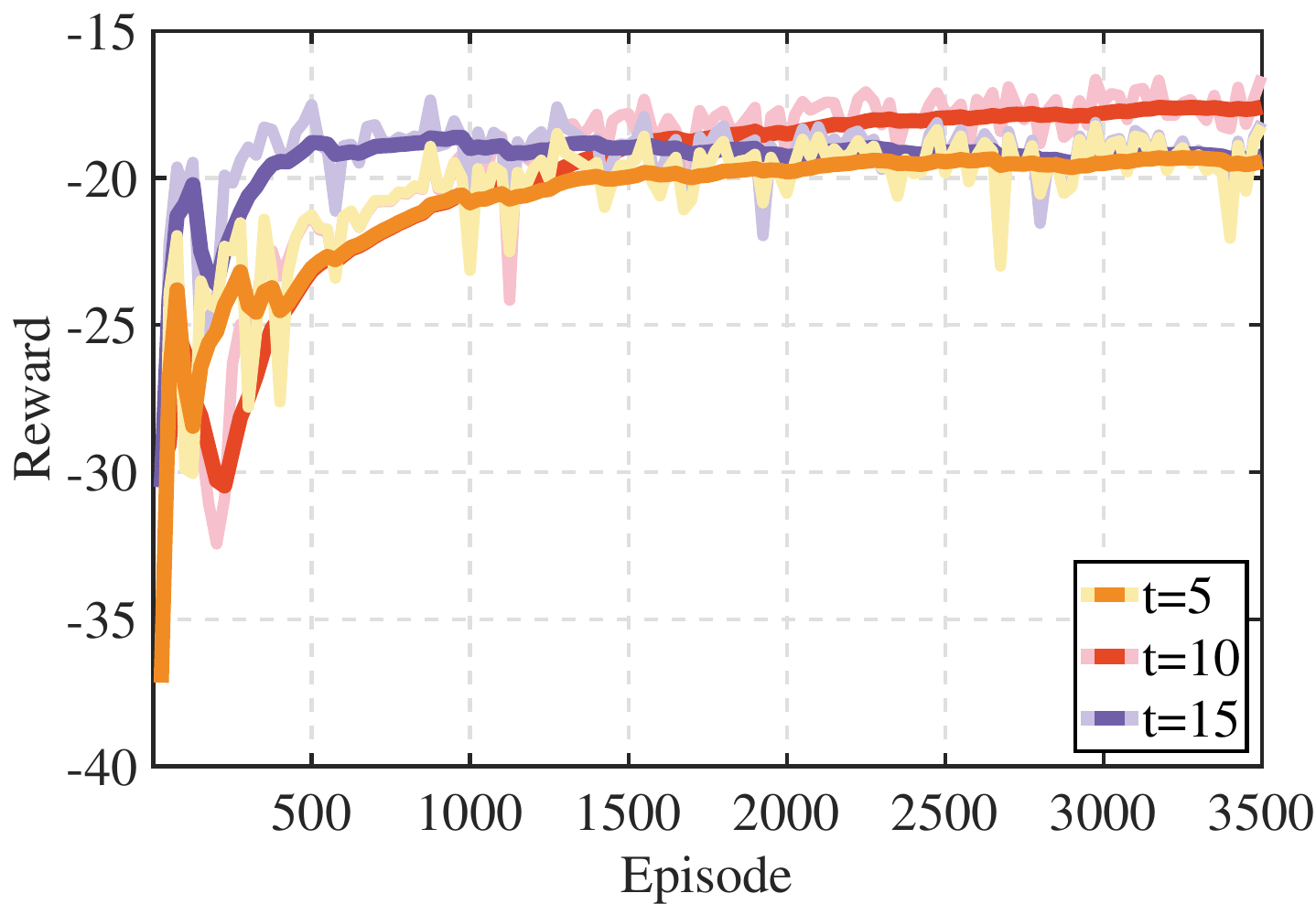}
    }\hfill
    \subfigure[Learning rate]{
        \includegraphics[width=0.3\linewidth]{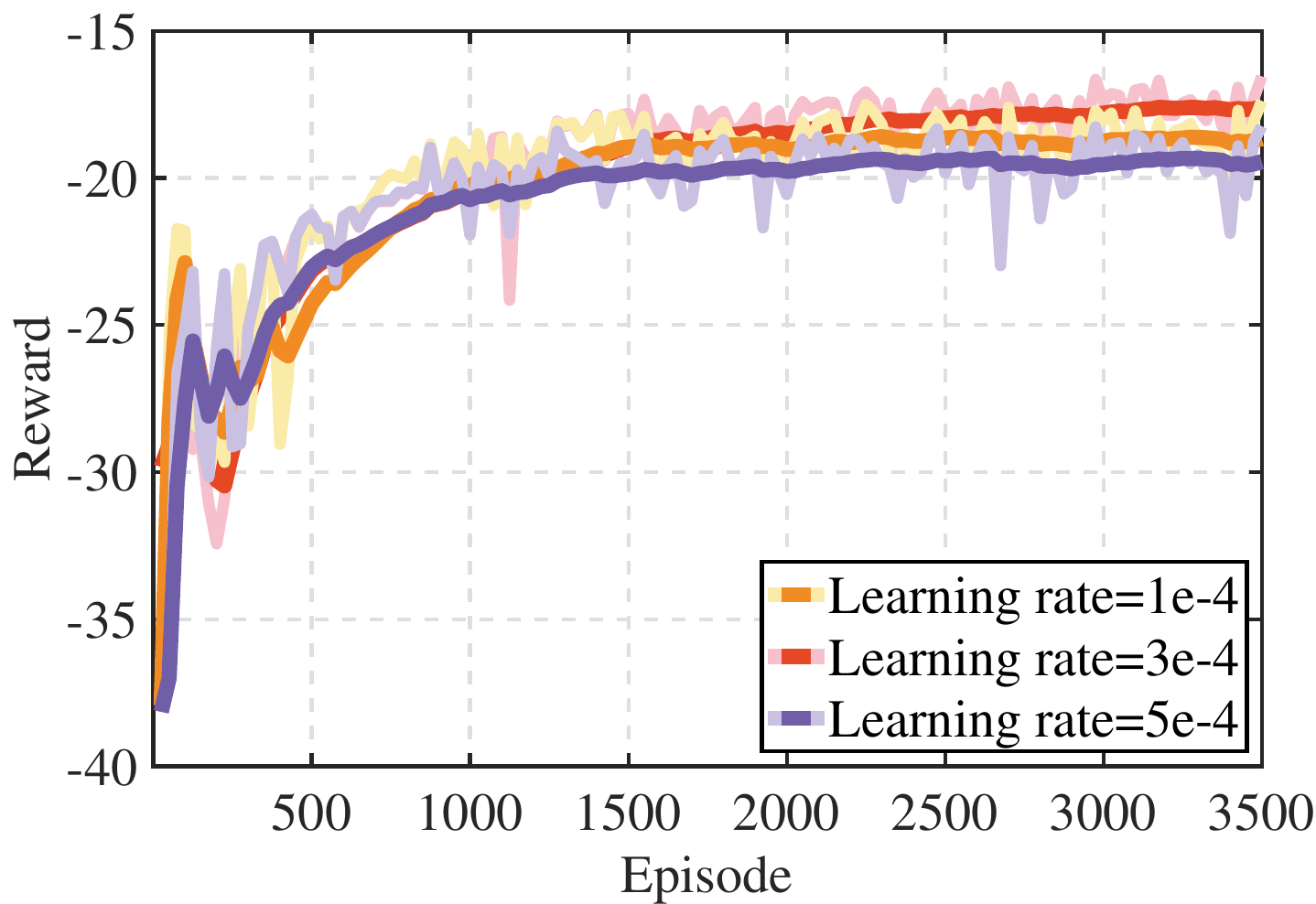}
    }\hfill
    \subfigure[Min-batch size]{
        \includegraphics[width=0.3\linewidth]{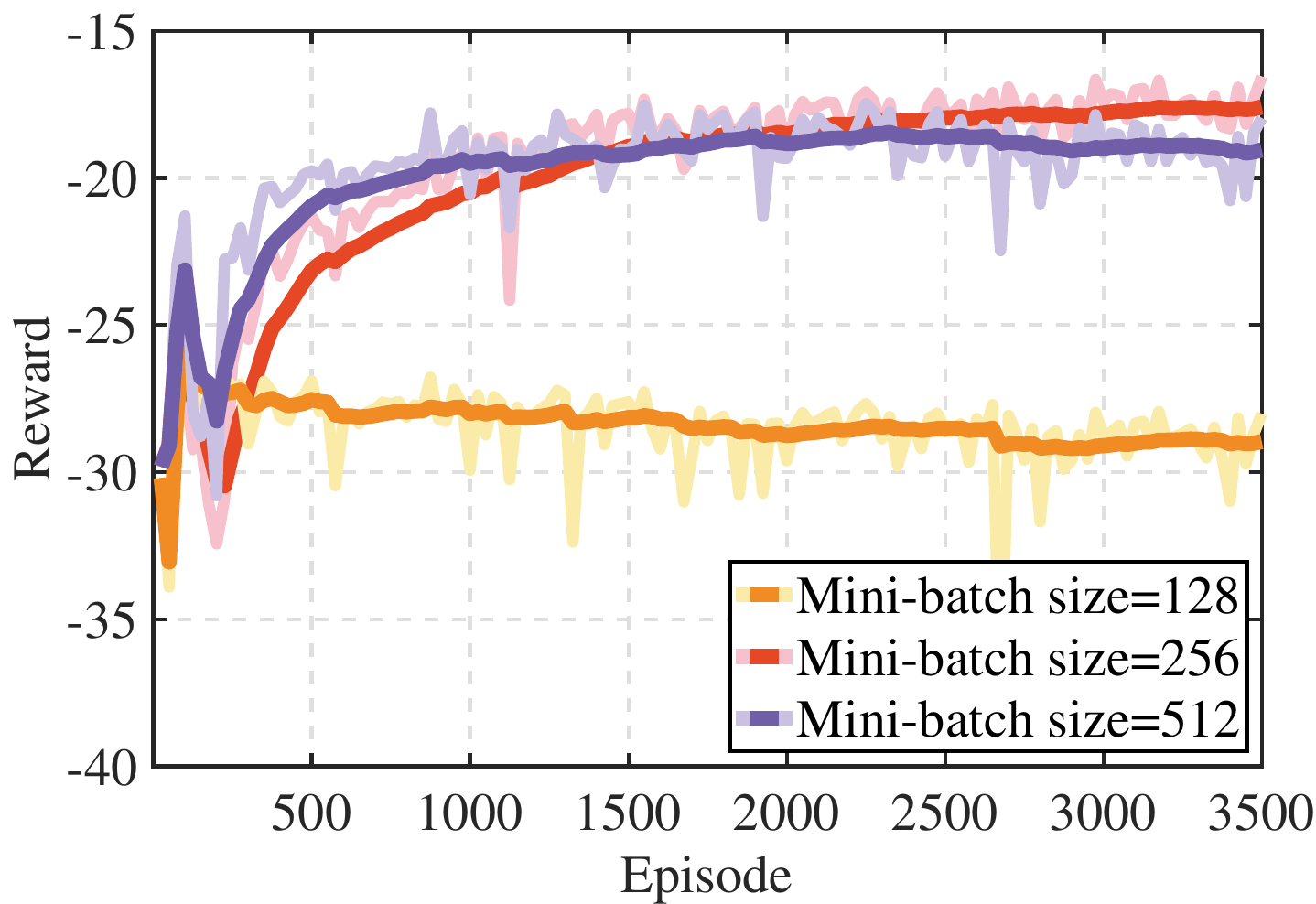}
    }
    \vspace{-1.2em}
    \caption{Reward comparison under different hyper-parameter settings of the HOOA approach. (a) Different diffusion steps. (b) Different learning rates. (c) Different min-batch sizes.}
    \label{fig_Reward}
    \vspace{-1.8em}
\end{figure*}

\vspace{-0.6em}
\subsubsection{Hyper-Parameter Sensitivity Analysis}
\label{sec:hyperparameter_sensitivity}
\vspace{-0.1em}
\par \textit{\textbf{Impact of Different Diffusion Steps.}} Fig.~\ref{fig_Reward}(a) shows the impact of different diffusion steps on the reward during the training process. As the diffusion step increases from $t=5$ to $t=10$, the reward improves more rapidly and converges to a higher level. When the diffusion step is further increased to $t=15$, the reward starts from a noticeably higher value and rises faster in the early episodes. This indicates that more denoising steps can yield better initial actions and accelerate early stage learning. However, the improvement in the later stage becomes marginal and the converged reward remains lower than the optimal setting of $t=10$. This is because an overly long denoising chain may introduce redundant refinement and additional sampling noise, which weakens the efficiency of policy updates and makes the training more prone to settling at a slightly inferior fixed point.

\par \textit{\textbf{Impact of Different Learning Rates.}} Fig.~\ref{fig_Reward}(b) compares the reward under different learning rates during the training process. Specifically, the three curves exhibit similar reward growth within the initial approximately 1000 episodes, while clear differences emerge in the later stage. The learning rate of $3\times10^{-4}$ continues to make steady gains and converges to the best final reward, which reveals that a moderate learning rate strikes a better trade-off between update stability and performance improvement. At a learning rate of $1\times10^{-4}$, overly small updates reduce learning efficiency and slow the correction of suboptimal actions, thus leading to a lower converged reward. When the learning rate increases to $5\times10^{-4}$, the reward improvement saturates earlier and ends up the worst, as overly aggressive updates can destabilize the value estimation and undermine stable fine-tuning in the later stage.

\par \textit{\textbf{Impact of Different Mini-Batch Sizes.}} Fig.~\ref{fig_Reward}(c) illustrates the impact of different mini-batch sizes on the reward during training. With a mini-batch size of $256$, the reward rises rapidly in the early stage, continues to improve steadily in the mid-to-late stage, and finally converges to the highest level with only minor fluctuations. Increasing the mini-batch size to $512$ leads to a strong early increase and competitive performance, but the subsequent improvement becomes weaker and the curve settles at a slightly lower level than $256$. This may be because overly large mini-batches can dampen gradient diversity and reduce update responsiveness, thereby making fine-grained refinement in the later stage less effective. In contrast, the mini-batch size of $128$ shows a distinctive pattern. After a brief initial rise, the reward drops back and then remains trapped at a low plateau. This indicates that small mini-batches introduce noisy gradients, which destabilize value learning and prevent the policy from escaping a suboptimal plateau.

\par Consequently, the analysis above validates the effectiveness of the hyper-parameter configuration adopted in the proposed HOOA. The selected diffusion steps, learning rate, and mini-batch size are crucial for ensuring robust training and strong final performance.

\vspace{-0.8em}
\subsubsection{Impact of System Settings}
\label{sec:impact_system_settings} 

\par
\begin{figure*}[!hbt]
    \centering
    \subfigure[]{
        \includegraphics[width=0.23\linewidth]{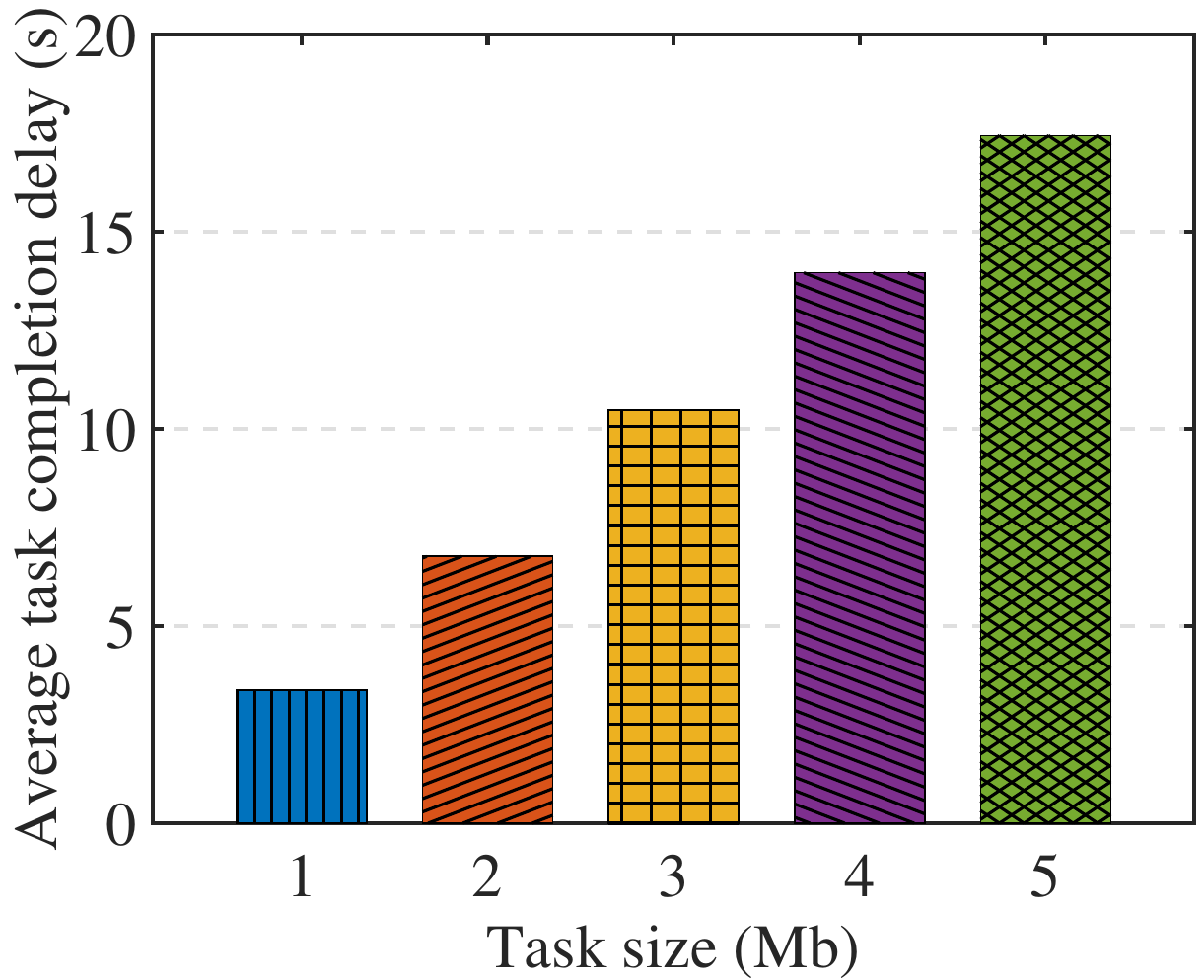}
    }\hfill
    \subfigure[]{
        \includegraphics[width=0.23\linewidth]{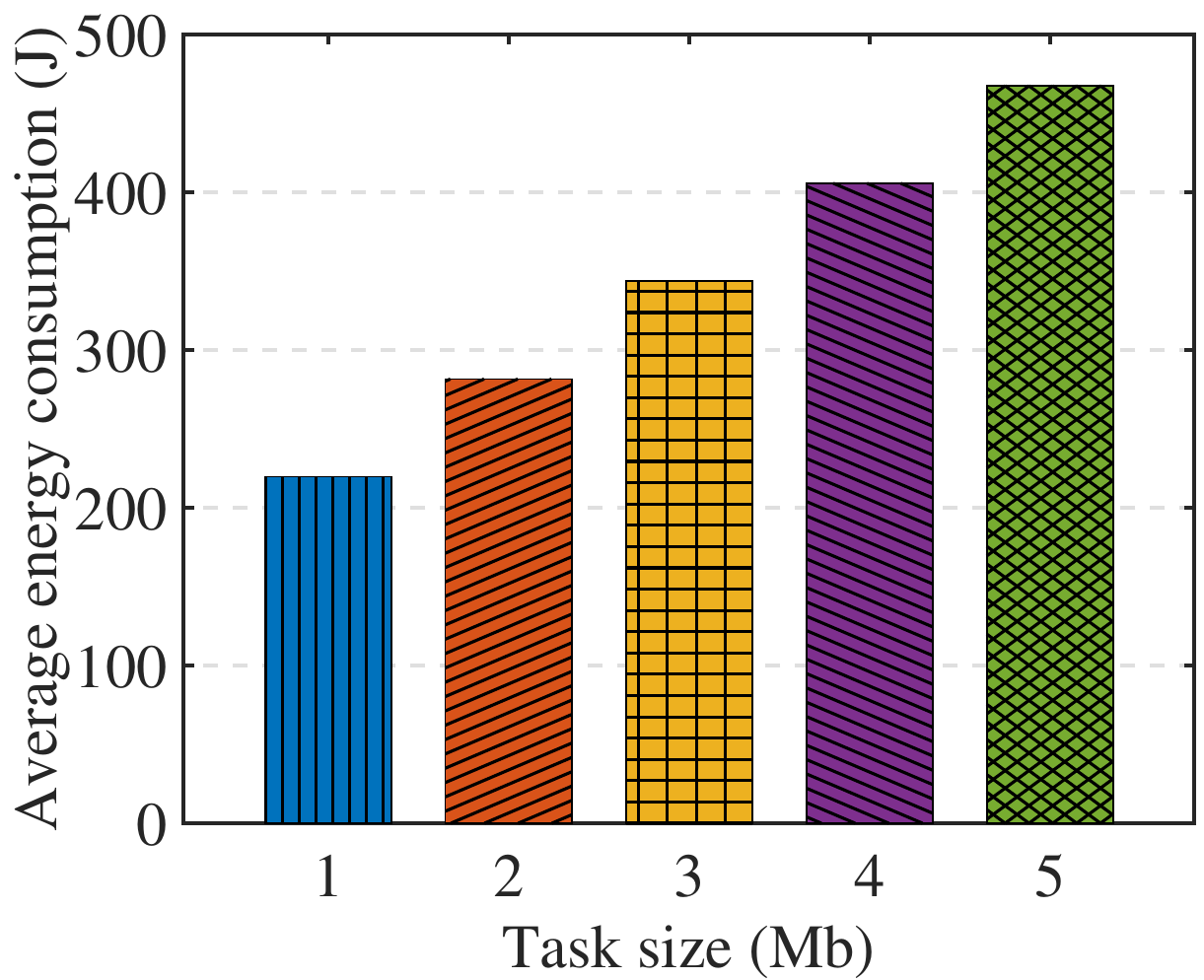}
    }\hfill
    \subfigure[]{
        \includegraphics[width=0.23\linewidth]{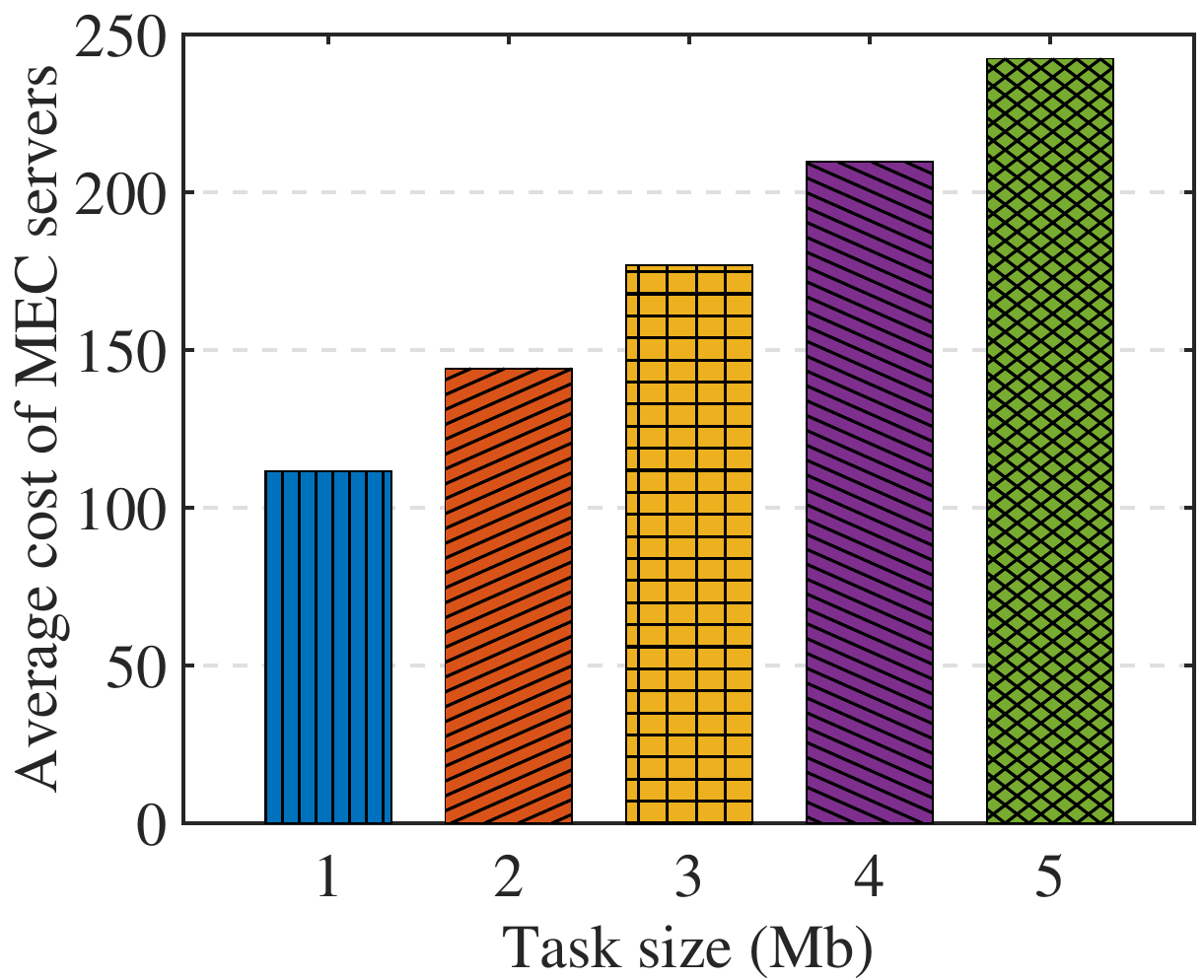}
    }\hfill
    \subfigure[]{
        \includegraphics[width=0.23\linewidth]{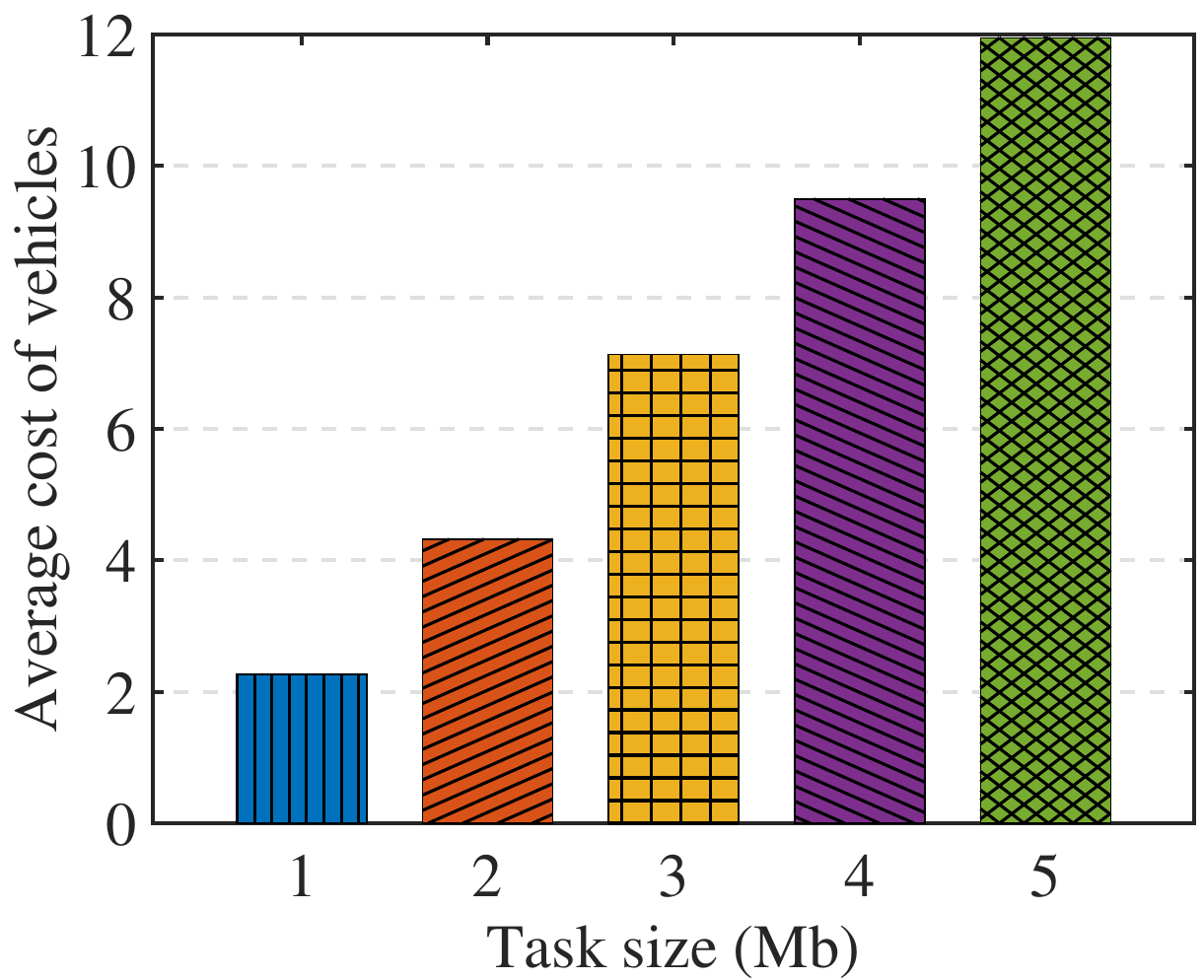}
    }
    \vspace{-1.2em}
    \caption{System performance of the HOOA approach with the task size. (a) Average task completion delay. (b) Average energy consumption. (c) Average cost of MEC servers. (d) Average cost of vehicles.}
    \label{fig_task_size}
    \vspace{-1.8em}
\end{figure*}

 \textit{\textbf{Impact of Task Size.}} Figs.~\ref{fig_task_size}(a), \ref{fig_task_size}(b), \ref{fig_task_size}(c), and \ref{fig_task_size}(d) present the impact of task size on the proposed HOOA in terms of average task completion delay, average energy consumption, average cost of MEC servers, and average cost of vehicles. It can be seen that as the task size increases from 1 Mb to 5 Mb, all the above metrics exhibit a clear increasing trend. Specifically, larger tasks introduce more data to be uploaded and impose heavier computation loads on MEC servers, which directly increases the task completion delay. Moreover, longer uplink transmissions, heavier computing workloads, and more proactive maneuvers by the UAV to maintain link quality and meet task deadlines all drive up the overall energy consumption. In addition, the cost of MEC servers accounts for not only the delay and energy consumed by MEC servers when executing offloaded tasks, but also the delay and energy incurred by vehicles during offloading when receiving MEC services. Meanwhile, the cost of vehicles is determined by task completion delay and vehicle-side energy consumption. Therefore, both metrics naturally increase as the task size grows.

\begin{figure*}[!hbt]
    \centering
    \subfigure[]{
        \includegraphics[width=0.23\linewidth]{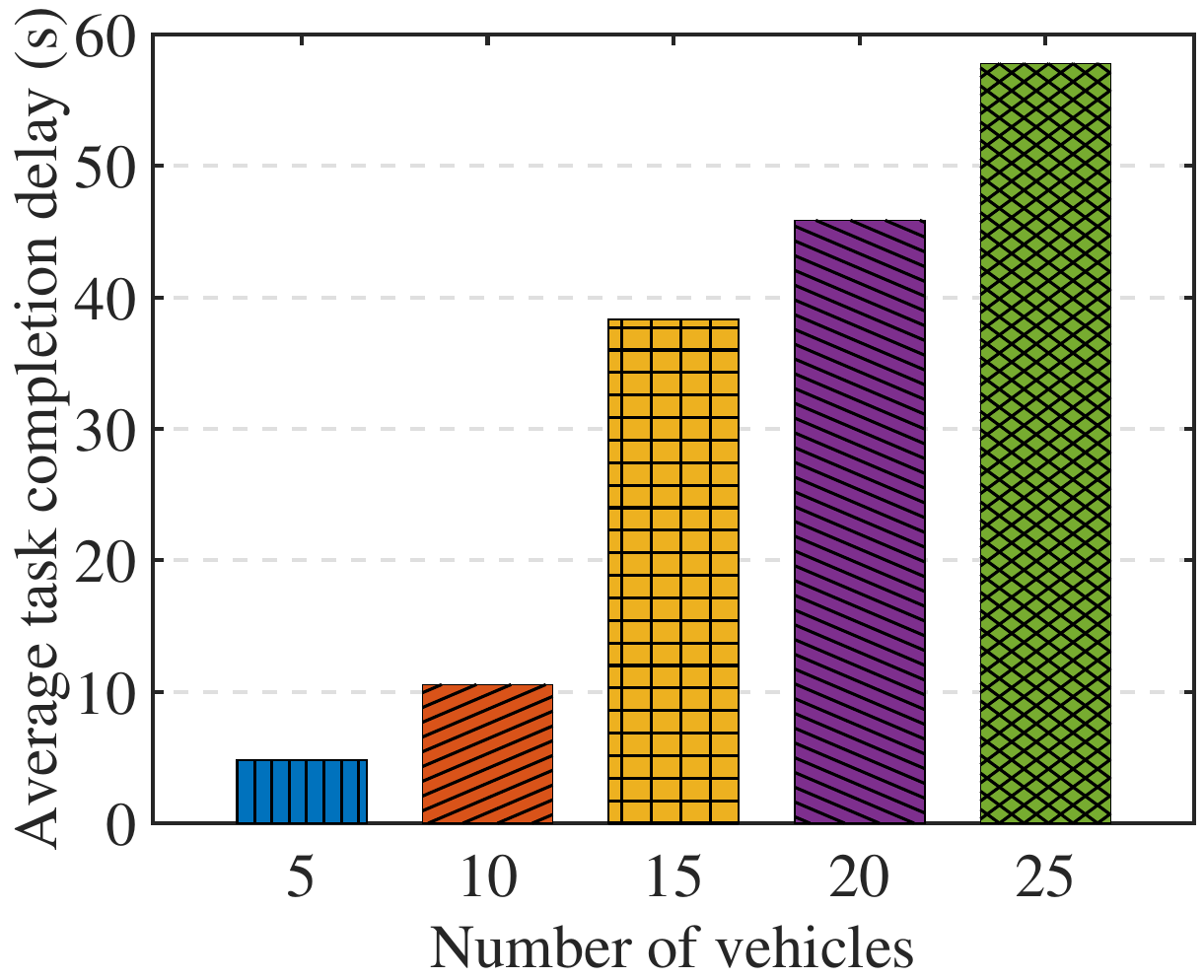}
    }\hfill
    \subfigure[]{
        \includegraphics[width=0.23\linewidth]{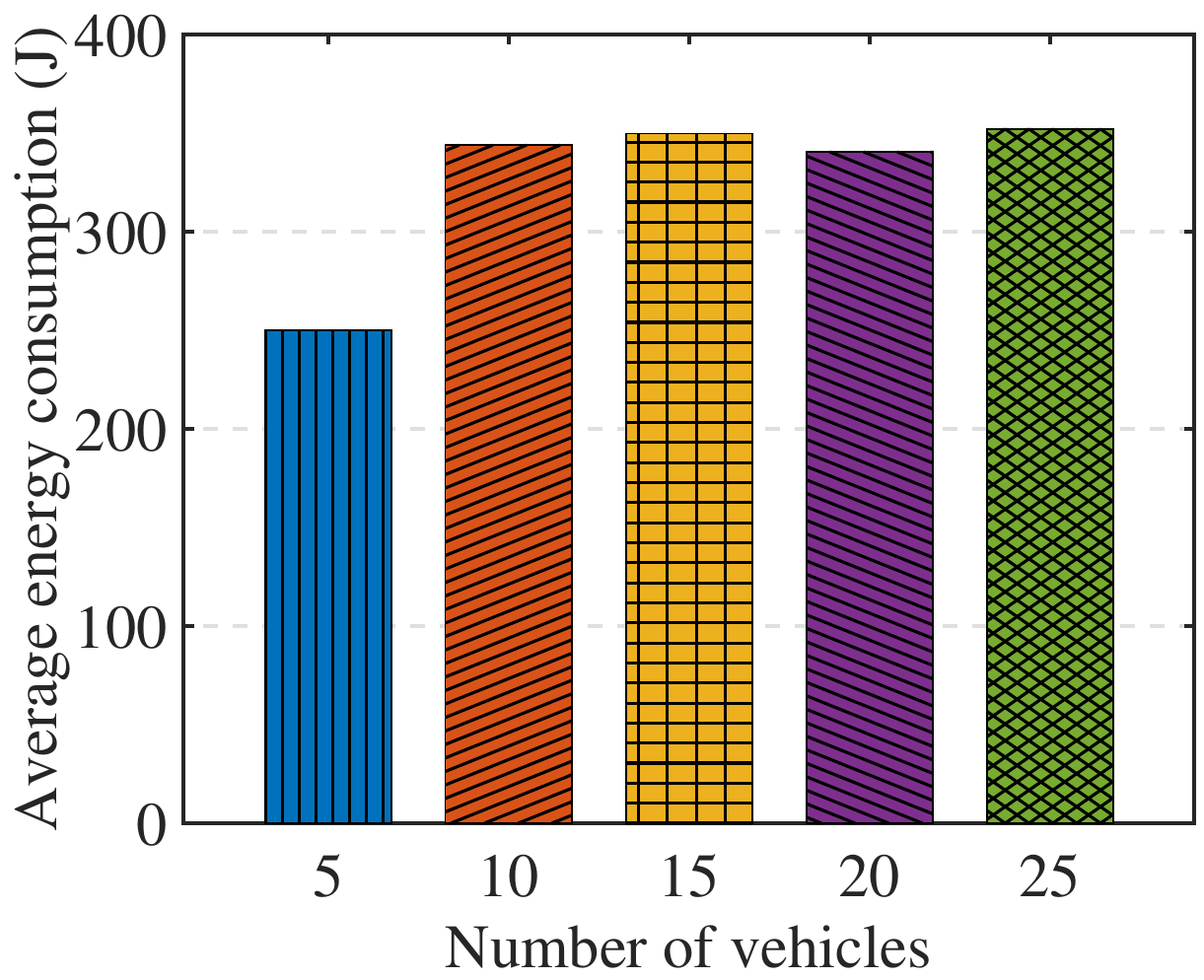}
    }\hfill
    \subfigure[]{
        \includegraphics[width=0.23\linewidth]{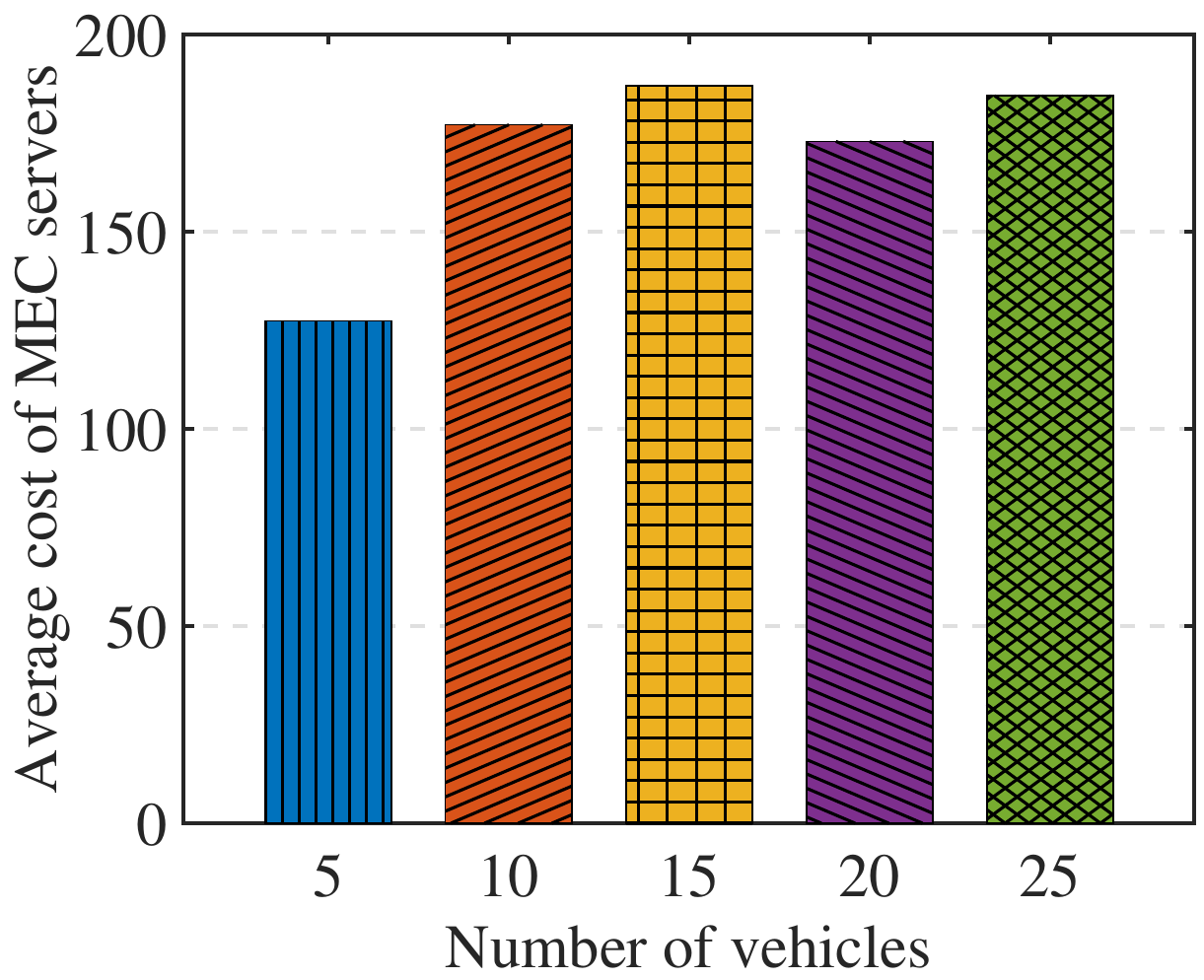}
    }\hfill
    \subfigure[]{
        \includegraphics[width=0.23\linewidth]{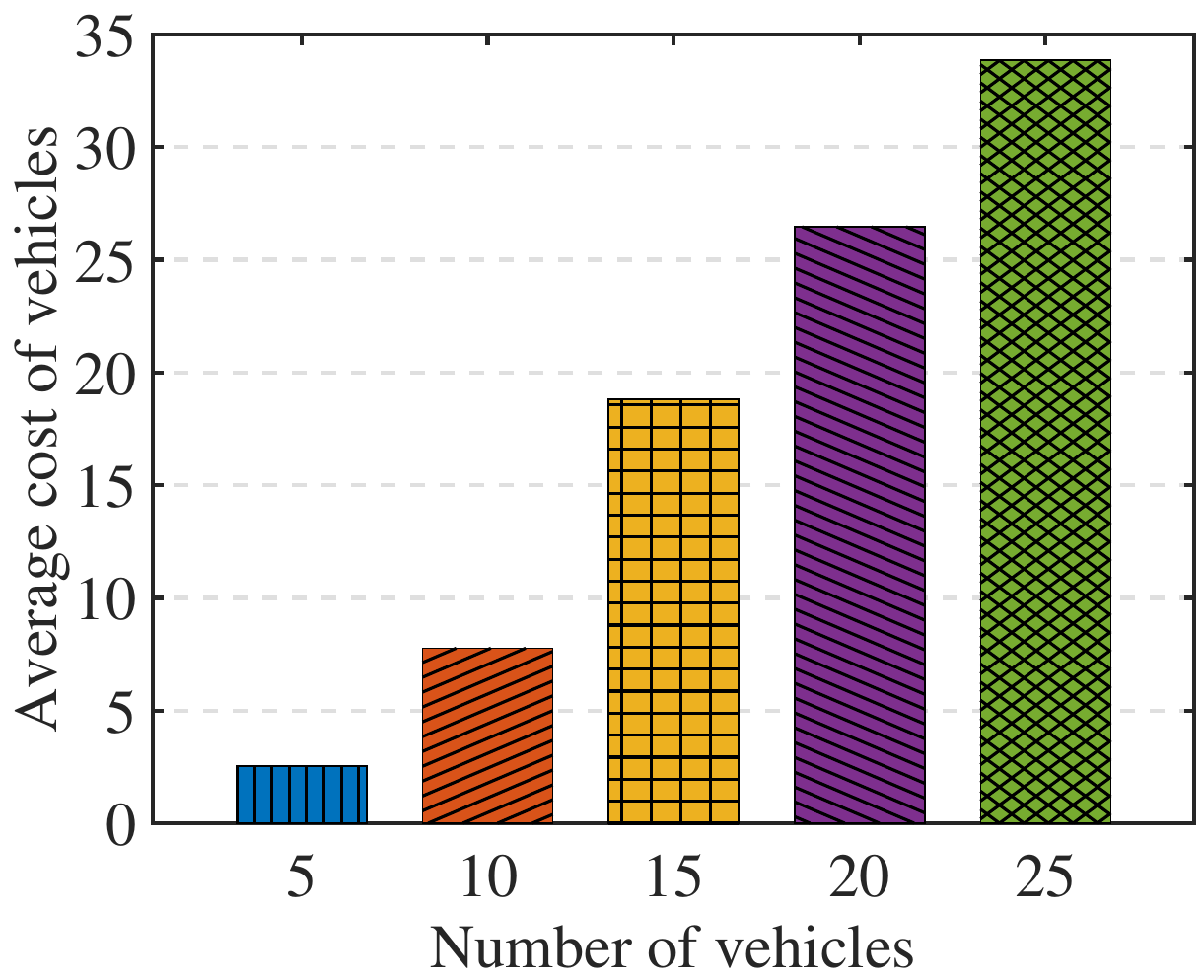}
    }
    \vspace{-1.2em}
    \caption{System performance of the HOOA approach with the number of vehicles. (a) Average task completion delay. (b) Average energy consumption. (c) Average cost of MEC servers. (d) Average cost of vehicles.}
    \label{fig_vehicles}
    \vspace{-1.8em}
\end{figure*}

\par \textit{\textbf{Impact of Number of Vehicles.}} Figs.~\ref{fig_vehicles}(a), \ref{fig_vehicles}(b), \ref{fig_vehicles}(c), and \ref{fig_vehicles}(d) evaluate the proposed HOOA with different numbers of vehicles in terms of average task completion delay, average energy consumption, average cost of MEC servers, and average cost of vehicles. As the number of vehicles increases from 5 to 25, the task completion delay grows sharply. This is because more vehicles generate concurrent tasks and intensify contention for wireless links and limited MEC resources, so that serving all vehicles efficiently in each slot becomes more difficult. A notable observation in Fig.~\ref{fig_vehicles}(b) is that the energy consumption increases significantly from 5 to 10 vehicles and then exhibits a saturation trend with minor oscillations. The reason is that, under moderate to high load, the MEC server capacity limits the number of tasks that can be offloaded or served in each slot. At the same time, HOOA adaptively shifts some tasks to local processing and tends to select smoother UAV maneuvers, thereby preventing the energy consumption from growing linearly. Similarly, due to the definitions of the cost of MEC servers and the cost of vehicles, both metrics are closely tied to task completion delay and energy consumption, and thus vary with the number of vehicles. 

\par In conclusion, the results under different task sizes and varying numbers of vehicles demonstrate that the proposed HOOA achieves strong robustness and scalability in dynamic vehicular network scenarios.
\vspace{-1em}
\section{Conclusion}
\label{sec:conclusion}
\par In this work, we have studied an IRS-enabled low-altitude MEC architecture for vehicular networks, where an aerial MEC server cooperates with a terrestrial MEC server and a hybrid IRS deployment (i.e., building-installed and UAV-carried IRSs) enhances air-ground connectivity under blockage. Based on this architecture, we have formulated the MOOP to minimize the task completion delay and energy consumption by jointly optimize task offloading, UAV trajectory control, IRS phase-shift configuration, and computation resource allocation. To efficiently solve this challenging problem, we have proposed the HOOA approach by reformulating MOOP as a Stackelberg game with follower-level and leader-level problems. Specifically, for the follower-level problem, we propose a many-to-one matching mechanism to generate feasible discrete decisions. For the leader-level problem, we leverage the GDMTD3 algorithm to enhance the action representation and exploration for continuous decisions, and further integrate a KKT-based method to reduce the action dimensionality. Extensive simulations validate that the proposed HOOA consistently outperforms benchmark approaches in terms of system performance, while also exhibiting faster convergence and more stable learning behavior compared with state-of-the-art DRL algorithms. Moreover, evaluations across different task sizes and varying numbers of vehicles confirm its strong robustness and scalability in dynamic vehicular scenarios.
\vspace{-1.5em}
\bibliographystyle{IEEEtran}
\bibliography{references.bib}

\end{document}